\documentclass[reprint,aps,prl,twocolumn,floatfix,hidelinks]{revtex4-2}

\usepackage{graphicx}
\usepackage{siunitx}
\usepackage{amsmath}
\usepackage{amssymb}
\usepackage{csquotes}
\usepackage{soul}
\usepackage{hyperref}
\usepackage[nolist]{acronym}
\usepackage{braket}
\newcommand{\Title}{Robust Quantum Memory in a Trapped-Ion Quantum Network Node}
\newcommand{\RamanWavelength}{\SI{402}{\nano\meter}}
\newcommand{\OopModeFrequency}{\SI{3.354}{\mega\hertz}}
\newcommand{\iSwapFidelity}{\num{0.913(3)}}
\newcommand{\iSwapSubspaceFidelity}{\num{0.977(7)}}
\newcommand{\ZeemanLifetime}{\SI{2}{\milli\second}}
\newcommand{\ZeemanFidelity}{\num{0.97(2)}}
\newcommand{\ClockFidelity}{\num{0.93(2)}}

\newcommand{\TenSecFidelity}{0.81(4)}
\newcommand{\ExcessHeating}{\num{9.3(9)e-4}}
\newcommand{\LightShiftPerAttempt}{\SI{8.8(6)}{\micro\radian}}
\newcommand{\IPHeatingRate}{\SI{2700}{\per\second}}
\newcommand{\OOPHeatingRate}{\SI{360}{\per\second}}

\newcommand{\Zeeman}{N}
\newcommand{\Stretch}{L}
\newcommand{\Clock}{M}

\newcommand{\ion}[2]{\mbox{$^{#2}$#1$^+$}}
\newcommand{\Ca}[1]{\ion{Ca}{#1}}

\newcommand{\Sr}[1]{\ion{Sr}{#1}}

\newcommand{\lev}[2]{{#1}_{#2}}
\newcommand{\fslev}[3]{{#1}_{#2},\,m_{J}\!=\!{#3}}  % m not subscript

\newcommand{\hfslevshort}[2]{F\!=\!{#1},\,m_{F}\!=\!{#2}}
\newcommand{\ionstate}[2]{{#2}_\mathrm{#1}}

\newcommand{\ish}{\mbox{$\sim$}\,}

\begin{document}
\title{\Title{}}

\author{P.~Drmota}
\email{peter.drmota@physics.ox.ac.uk}
\author{D.~Main}
\author{D.~P.~Nadlinger}
\author{B.~C.~Nichol}
\author{M.~A.~Weber}
\author{E.~M.~Ainley}
\author{A.~Agrawal}
\author{R.~Srinivas}
\author{G.~Araneda}
\author{C.~J.~Ballance}
\author{D.~M.~Lucas}
\affiliation{Department of Physics, University of Oxford, Clarendon Laboratory, Parks Road, Oxford OX1 3PU, United Kingdom}

% PRL abstract (600 chars):
\begin{abstract}\noindent
We integrate a long-lived memory qubit into a mixed-species trapped-ion quantum network node.
Ion-photon entanglement first generated with a network qubit in \Sr{88} is transferred to \Ca{43} with \iSwapSubspaceFidelity{} fidelity, and mapped to a robust memory qubit.
We then entangle the network qubit with a second photon, without affecting the memory qubit.
We perform quantum state tomography to show that the fidelity of ion-photon entanglement decays \ish\num{70} times slower on the memory qubit.
Dynamical decoupling further extends the storage duration; we measure an ion-photon entanglement fidelity of \TenSecFidelity{} after \SI{10}{\second}.
\end{abstract}

\maketitle
\noindent
Quantum networks have the potential to revolutionize the way we distribute and process information~\cite{kimble_quantum_2008}.
They have applications in cryptography~\cite{bennett_quantum_1984, ekert_quantum_1991}, quantum computing~\cite{bennett_teleporting_1993, hucul_modular_2015}, and metrology~\cite{komar_quantum_2014}, and can provide insights into the nature of entanglement~\cite{matsukevich_bell_2008, hensen_loophole-free_2015}.
Photonic interfaces are essential for such networks, enabling two remote stationary qubits to exchange quantum information using entanglement swapping~\cite{northup_quantum_2014}.
Elementary quantum networks have been realized with diamond nitrogen-vacancy centers~\cite{hensen_loophole-free_2015, hermans_qubit_2022}, photons~\cite{jennewein_quantum_2000, yin_quantum_2012}, neutral atoms~\cite{matsukevich_entanglement_2006, nolleke_efficient_2013, zhang_device-independent_2022}, solid-state systems~\cite{magnard_microwave_2020}, and trapped ions~\cite{moehring_entanglement_2007, matsukevich_bell_2008, olmschenk_quantum_2009, pironio_random_2010, stephenson_high-rate_2020, nadlinger_experimental_2022, nichol_elementary_2022, krutyanskiy_entanglement_2022, krutyanskiy_telecom-wavelength_2022}. % We love trapped ions.

% Why are trapped ions great?
Trapped ions provide qubits with exceptionally long coherence times, which can be initialized, manipulated, entangled, and read out with high fidelity~\cite{wang_single_2021, christensen_high-fidelity_2020, harty_high-fidelity_2014, srinivas_high-fidelity_2021, gaebler_high-fidelity_2016, clark_high-fidelity_2021}.
Moreover, trapped ions readily interact with optical fields, providing a natural interface between their electronic state -- the stationary quantum memory -- and photons -- the \enquote{flying} quantum information carrier~\cite{blinov_observation_2004}.
Pairs of trapped-ion network nodes comprising one qubit of a single species have been connected by a photonic link and used to perform Bell tests~\cite{matsukevich_bell_2008}, state teleportation~\cite{olmschenk_quantum_2009}, random number generation~\cite{pironio_random_2010}, quantum key distribution~\cite{nadlinger_experimental_2022}, and frequency comparisons~\cite{nichol_elementary_2022}.
Trapped ion systems have also demonstrated state-of-the-art single- and two-qubit gate fidelities, but integrating these within a quantum network node remains a challenge since an ion species suitable for quantum communication does not necessarily also provide a good memory qubit with sufficient isolation from network activity.
Atomic species such as \ion{Ba}{133} or \ion{Yb}{171} have been proposed to circumvent this issue \cite{christensen_high-fidelity_2020, yang_realizing_2022}; however, the development of the required experimental techniques is still ongoing.
Alternatively, it is possible for each role to be fulfilled by a different species~\cite{inlek_multispecies_2017}.
In addition, using multiple atomic species has advantages for minimizing crosstalk during mid-circuit measurements and cooling~\cite{negnevitsky_repeated_2018}.
%The coherent transfer of information from the photon onto the memory qubit is a requirement for these quantum networks, but has not been demonstrated yet.

In this Letter, we demonstrate a trapped-ion quantum network node in which entanglement between a network qubit and a photon is created and coherently transferred onto a memory qubit for storage, while the network qubit is entangled with a second photon.
Due to its simple level structure, \Sr{88} is ideally suited for our \ac{IPE} scheme~\cite{stephenson_high-rate_2020}, whereas the hyperfine structure of \Ca{43} provides a long-lived memory qubit~\cite{lucas_long-lived_2007}.
While both \ac{IPE} and local mixed-species entangling gates have been demonstrated independently~\cite{inlek_multispecies_2017}, this is the first experiment in which these capabilities are combined.
% environmental noise, or more specifically B-field?
Furthermore, we show that the memory qubit in \Ca{43} is robust to environmental noise as well as to concurrent addressing of \Sr{88} for the generation of \ac{IPE}.
Finally, sympathetic cooling of the ion pair using \Sr{88} between rounds of entanglement generation enables continued operation even in the presence of heating.

For this experiment, we load a \Sr{88}-\Ca{43} crystal with controlled order into a surface-electrode Paul trap at room temperature \footnote{Sandia National Laboratories HOA2.}.
Each experimental sequence begins with cooling~\cite{supplement}, reducing the temperature of the axial \ac{OOP} and \ac{IP} motion to $\bar{n}_{\mathrm{oop}}\simeq\num{0.3}$ and $\bar{n}_{\mathrm{ip}}\simeq\num{3}$, respectively.
The cooling sequence was empirically optimized for the high heating rates observed, namely $\dot{\bar{n}}_{\mathrm{oop}}\simeq\OOPHeatingRate{}$ at $\omega_\mathrm{oop}/(2\pi)=\OopModeFrequency{}$ and $\dot{\bar{n}}_{\mathrm{ip}}\simeq\IPHeatingRate{}$ at $\omega_\mathrm{ip}/(2\pi) = \SI{1.705}{\mega\hertz}$.
\begin{figure}[hb]
    \vspace{-0.3cm}
    \includegraphics{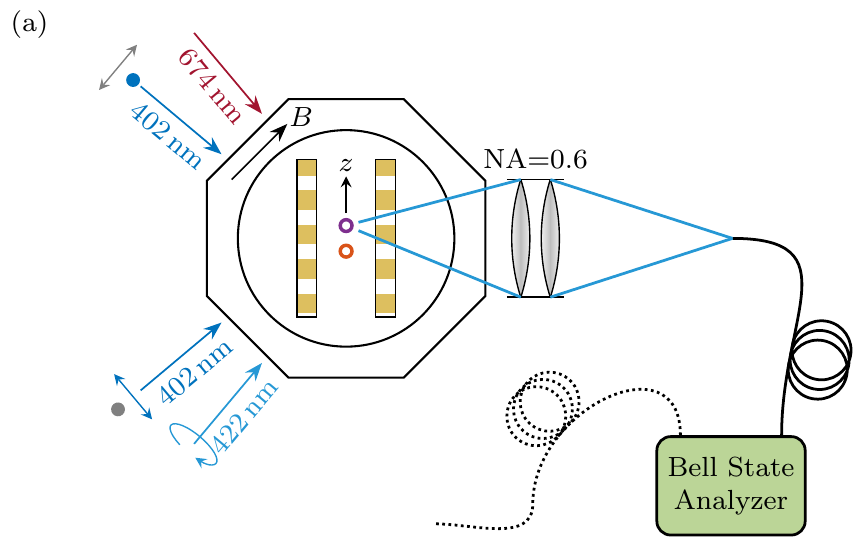}
    \includegraphics{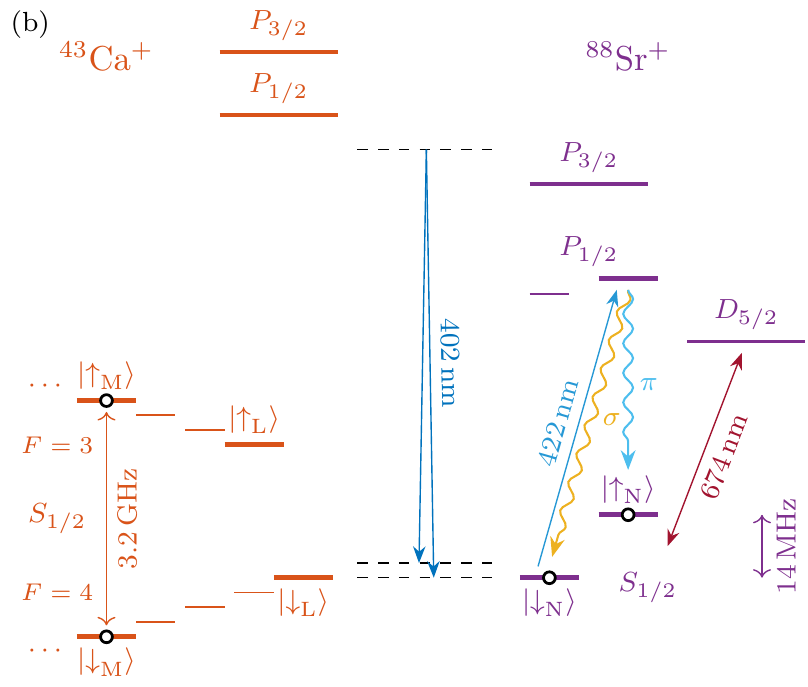}
    \caption{(a) Overview of the apparatus.
        We show the laser beam geometry; within the plane of the trap surface, the magnetic field $B$ is oriented at \ang{45} to the trap axis $z$. Perpendicular to this plane, the $\mathrm{NA}=0.6$ lens collects single photons from a \Sr{88} ion (violet circle).
        Single photons are coupled into a single-mode fiber that is connected to a Bell state analyzer.
        Here, only one network node is connected; the same device can herald remote entanglement with a second, identical node~\cite{stephenson_high-rate_2020}.
        The state of \Sr{88} can be mapped onto a co-trapped \Ca{43} ion (orange circle).
        (b) Level structure of \Sr{88} (violet) and \Ca{43} (orange), not to scale.
        The memory qubit comprises the $m_F=0$ states in the \Ca{43} $\lev{S}{1/2}$ manifold.
        Raman lasers (blue arrows, \SI{422}{\nano\meter}) are used to drive mixed-species entangling gates and transitions between \Ca{43} hyperfine ground states.
        A $\sigma^{+}$-polarized laser pulse excites the $\lev{S}{1/2} \leftrightarrow \lev{P}{1/2}$ transition in \Sr{88} to generate a single photon whose polarization (see $\sigma$ and $\pi$ decay channels) is entangled with the state of the ion.
        A narrow-linewidth laser (red arrow, \SI{674}{\nano\meter}) is used to manipulate the \Sr{88} qubit via the quadrupole transition.
    }
    \label{fig:ions}
\end{figure}
To produce single photons, \Sr{88} is excited to the $\ket{\fslev{P}{1/2}{+1/2}}$ state by a \ish\SI{10}{\pico\second} laser pulse.
This short-lived excited state decays with probability \num{0.95} into the $\lev{S}{1/2}$ manifold via emission of a photon at \SI{422}{\nano\meter} whose polarization is entangled with the spin state of the ion.
The photon emission is imaged by an $\mathrm{NA}=0.6$ objective onto a single-mode optical fiber [Fig.~\ref{fig:ions}(a)], which acts as a spatial mode filter, maximizing the entangled fraction by suppressing polarization mixing.
The ion-photon state can then be described by the maximally entangled Bell state
\begin{equation*}
  \ket{\psi} = \tfrac{1}{\sqrt{2}}\, \big(\ket{\ionstate{\Zeeman{}}{\downarrow}} \otimes \ket{H} + \ket{\ionstate{\Zeeman{}}{\uparrow}} \otimes \ket{V} \!\big) \ ,
  \label{E:ionphoton}
\end{equation*}
where $\ket{H}$ and $\ket{V}$ are orthogonal linear polarization states of the photon, and $\ket{\ionstate{\Zeeman{}}{\downarrow}}$ and $\ket{\ionstate{\Zeeman{}}{\uparrow}}$ denote the network qubit states in the Zeeman ground state manifold of \Sr{88} [Fig.~\ref{fig:ions}(b)].
To analyze the polarization state of the photon, we employ polarizing beamsplitters and avalanche photodiodes, which are part of the same photonic Bell state analyzer used to herald remote entanglement between two network nodes~\cite{stephenson_high-rate_2020}.
The pulsed excitation sequence is repeated in a loop at an attempt rate of \SI{1}{\mega\hertz} until a photon is detected.
The polarization measurement basis is set at the beginning of a sequence using motorized waveplates.
Qubit manipulation of \Sr{88} is performed on the \SI{674}{\nano\meter} quadrupole transition, which is also used for electron shelving readout.

The second ion species, \Ca{43}, is chosen for its excellent coherence properties and the high level of control achieved in previous experiments~\cite{home_memory_2009, harty_high-fidelity_2014, ballance_high-fidelity_2016, sepiol_probing_2019}.
Furthermore, the mass ratio between \Ca{43} and \Sr{88} is reasonably favorable for sympathetic cooling~\cite{kielpinski_sympathetic_2000}, and the electronic level structure facilitates mixed-species gates~\cite{hughes_benchmarking_2020}.
For state preparation, polarized \SI{397}{\nano\meter} light optically pumps population into $\ket{\ionstate{\Stretch{}}{\downarrow}}$.
A pair of co-propagating Raman laser beams at $\lambda_{\mathrm{R}}=\RamanWavelength{}$ is used to manipulate states within the ground state manifold.
For readout, population is shelved using a pulse sequence of \SI{393}{\nano\meter} and \SI{850}{\nano\meter} light~\cite{myerson_high-fidelity_2008}.
At a magnetic field of \SI{0.5}{\milli\tesla}, the frequency of the memory qubit transition depends weakly on the magnetic field magnitude, with a sensitivity of \SI{122}{\kilo\hertz\per\milli\tesla}.
Compared to the sensitivity of the \Sr{88} Zeeman qubit of \SI{28}{\mega\hertz\per\milli\tesla}, the memory qubit is significantly more resilient to magnetic field noise.
In addition, the magnetic field at the position of the ions is actively stabilized using feedback and feedforward currents to the $\SI{10}{\nano\tesla}$ level, calibrated over the \SI{50}{\hertz} line cycle using \Sr{88} as a magnetic field probe~\cite{main_magnetic_2020}.

To swap information from \Sr{88} to \Ca{43}, we perform mixed-species $\hat{\sigma}_z \otimes \hat{\sigma}_z$ geometric phase gates using the state-dependent light shift force generated by a single pair of \ish\SI{20}{\milli\watt} Raman laser beams at \RamanWavelength{}.
Only one pair is required to drive both species thanks to their compatible electronic level structure~\cite{hughes_benchmarking_2020} [Fig.~\ref{fig:ions}(b)].
The main advantage of this scheme over Cirac-Zoller and M{\o}lmer-S{\o}rensen gates, which have previously been explored in this context~\cite{inlek_multispecies_2017, bruzewicz_dual-species_2019}, is its robustness to qubit frequency shifts.
%The latter is of particular significance for \ac{IPE} where the state phase is set by the detection time of the photon, which is fundamentally random.
The Raman beams are aligned to address the \ac{OOP} axial motion of the two-ion crystal~\cite{hughes_benchmarking_2020}.
For maximum gate efficiency on this mode, the ion spacing is set to a half-integer multiple of the effective wavelength $\lambda_{\mathrm{R}}/\sqrt{2}$.
As the memory qubit in \Ca{43} does not experience a light shift, this interaction is performed on the logic qubit \Stretch{} instead.
First-order Walsh modulation compensates for the light shift imbalance between the two species.
With the available laser power, a detuning of $\delta/(2\pi) = \SI{34}{\kilo\hertz}$ from the \ac{OOP} mode achieves a gate duration of $\approx\SI{60}{\micro\second}$ while minimizing off-resonant excitation of the \ac{IP} mode~\cite{supplement}.
Charging of the trap due to the Raman laser beams is automatically compensated every \ish\SI{5}{\minute} using the method described in Ref.~\cite{nadlinger_micromotion_2021}.

\begin{figure}[htb]
    \includegraphics{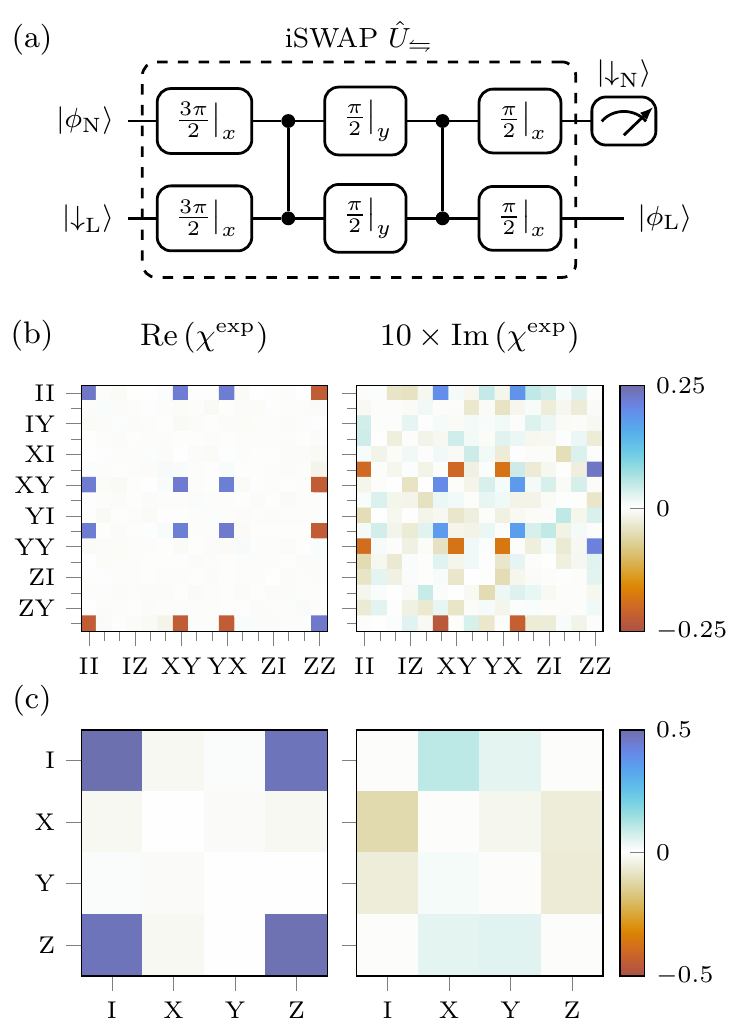}
    \caption{(a) The iSWAP circuit used to map the network qubit state from \Sr{88} to the logic qubit in \Ca{43}.
    (b) The Choi matrix reconstructed from process tomography of the iSWAP gate before error detection indicates a process fidelity of \iSwapFidelity{}.
    (c) Initializing the logic qubit in $\ket{\ionstate{\Stretch{}}{\downarrow}}$ and rejecting errors flagged by the measurement on the network qubit, the fidelity of this conditional process is \iSwapSubspaceFidelity{}.
    }
    \label{fig:iswap}
\end{figure}
The state of the network qubit in \Sr{88} is coherently swapped onto the logic qubit using an iSWAP gate, which is implemented by two $\hat{\sigma}_z \otimes \hat{\sigma}_z$ interactions and single-qubit rotations [circuit shown in Fig.~\ref{fig:iswap}(a)].
We use the iSWAP, as opposed to a full SWAP, since the initial state of \Ca{43} is known to be prepared in $\ket{\ionstate{\Stretch{}}{\downarrow}}$.
The ideal iSWAP performs the mapping $\ket{\ionstate{\Zeeman{}}{\phi}} \otimes \ket{\ionstate{\Stretch{}}{\downarrow}} \mapsto \ket{\ionstate{\Zeeman{}}{\downarrow}} \otimes \ket{\ionstate{\Stretch{}}{\phi}}$, where $\ket{\phi}$ is the arbitrary state to be swapped to the logic qubit, leaving the network qubit in the $\ket{\ionstate{\Zeeman{}}{\downarrow}}$ state.
Experimental imperfections leading to deviations from the target subspace $\ket{\ionstate{\Zeeman{}}{\downarrow}}\bra{\ionstate{\Zeeman{}}{\downarrow}} \otimes \hat{1}_\mathrm{\Stretch{}}$ are detected via a mid-circuit measurement on the network qubit~\cite{stas_robust_2022}.
We characterize the iSWAP operation independently using process tomography~\cite{rehacek_diluted_2007, anis_maximum-likelihood_2012} to reconstruct the Choi process matrix $\chi^\mathrm{exp}$ and calculate the process fidelity $\mathcal{F}_\mathrm{p} = \mathrm{Tr}(\chi^\mathrm{id}\chi^\mathrm{exp})$ with respect to the ideal process $\chi^\mathrm{id}$, yielding $\mathcal{F}_\mathrm{p} = \iSwapFidelity{}$ [Fig.~\ref{fig:iswap}(b)].
Considering only the mapping from the subspace $\hat{1}_\mathrm{\Zeeman{}} \otimes \ket{\ionstate{\Stretch{}}{\downarrow}}\bra{\ionstate{\Stretch{}}{\downarrow}}$ where the logic qubit is prepared in  $\ket{\ionstate{\Stretch{}}{\downarrow}}$ to the subspace $\ket{\ionstate{\Zeeman{}}{\downarrow}}\bra{\ionstate{\Zeeman{}}{\downarrow}} \otimes \hat{1}_\mathrm{\Stretch{}}$ where the network qubit is measured in $\ket{\ionstate{\Zeeman{}}{\downarrow}}$, the process fidelity is \iSwapSubspaceFidelity{} with respect to the ideal process $\ket{\ionstate{\Zeeman{}}{\phi}} \otimes \ket{\ionstate{\Stretch{}}{\downarrow}} \mapsto \ket{\ionstate{\Zeeman{}}{\downarrow}} \otimes \ket{\ionstate{\Stretch{}}{\phi}}$ [Fig.~\ref{fig:iswap}(c)].

\begin{figure*}[hbt]
    \includegraphics{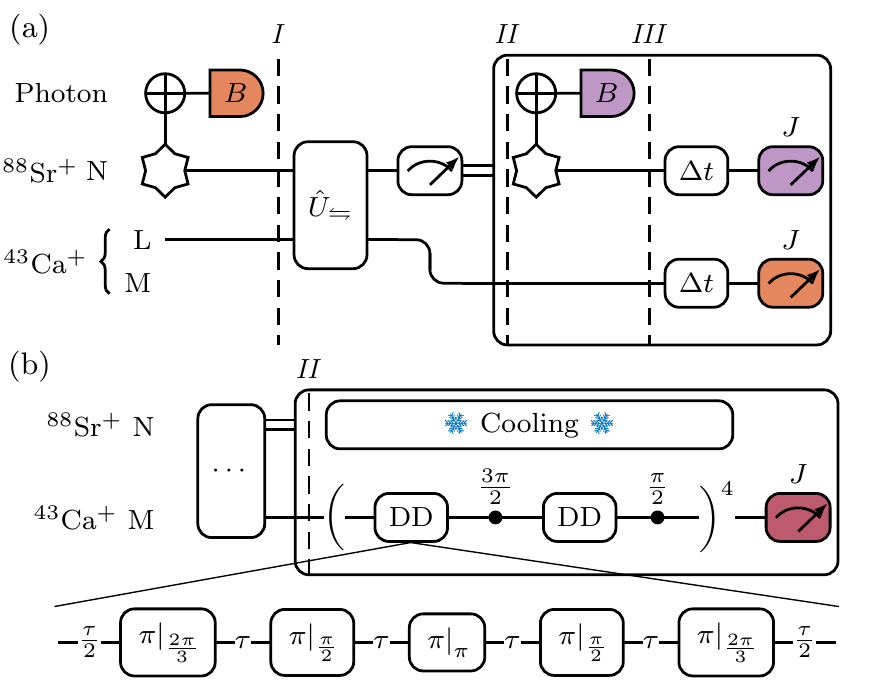}
    \includegraphics{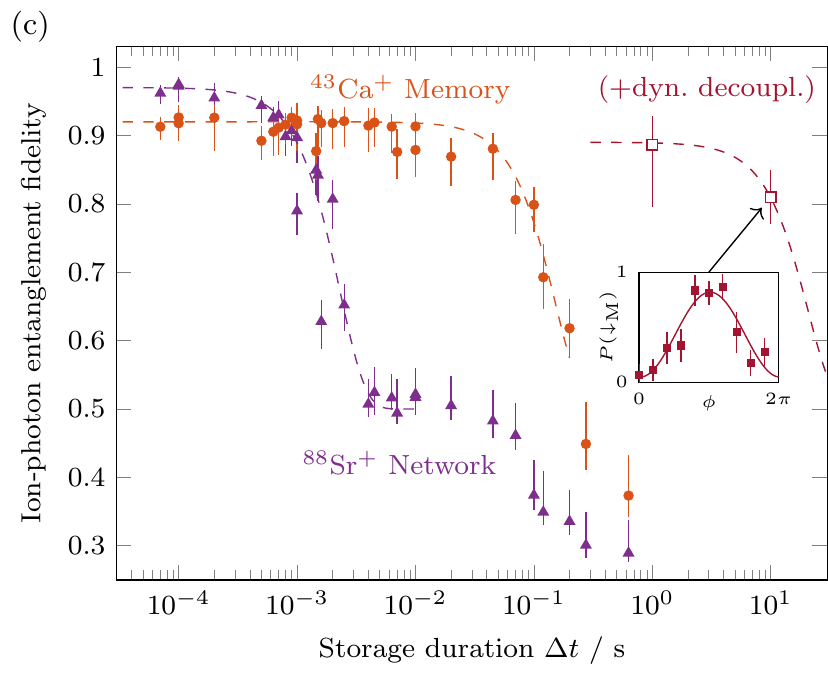}
    \caption{(a-b) Experimental sequences to probe the memory properties of the network node.
        Delayed measurements from a complete set of bases $B\otimes J$ are used to tomographically reconstruct the density matrices of the ion-photon states.
        If the mid-circuit measurement detects errors in the iSWAP gate, the sequence is immediately restarted.
        (a) A second photon is generated after transferring the state entangled with the first photon to the memory qubit.
        (b) After transferring \ac{IPE} from the network qubit to the memory qubit, \Sr{88} is used to sympathetically cool \Ca{43}.
        Dynamical decoupling and ion transport are used to extend the memory coherence time during the storage period.
        (c) The fidelities of the ion-photon states with respect to the closest maximally entangled state~\cite{supplement} are calculated from the density matrix obtained from maximum likelihood estimation~\cite{supplement} and averaged over all four photon detectors.
        Error bars span the \SI{95}{\percent} confidence interval obtained from nonparametric bootstrapping~\cite{supplement}.
        Dashed curves show Gaussian decay models to guide the eye.
        Square symbols indicate the fidelity with dynamical decoupling, ion transport, and sympathetic cooling using \Sr{88} during the storage time.
        At \SI{10}{\second}, only the populations and the parity were measured to infer the fidelity (see inset for the signal correlated with one photon detector, versus varying memory qubit rotation angle $\phi$).
    }
    \label{fig:results}
\end{figure*}

The iSWAP operation enables the transfer of \ac{IPE} from the network qubit in \Sr{88} to the memory qubit in \Ca{43}, so that \ac{IPE} can be created a second time using \Sr{88}.
To probe the memory properties of the integrated system of entangled photons and ions, we perform tomography on both ion-photon states in parallel after a variable storage duration.
For this, we initialize $\ket{\ionstate{\Stretch{}}{\downarrow}} \otimes \ket{\ionstate{\Zeeman{}}{\downarrow}}$ and execute the attempt loop until a single photon is detected [point~$\mathit{I}$ in Fig.~\ref{fig:results}(a)].
Subsequently, we swap the network qubit state to the logic qubit, and further to the memory qubit \Clock{} for storage~\cite{supplement}.
If the \SI{130}{\micro\second} mid-circuit measurement on the network qubit indicates a success [point~$\mathit{II}$ in Fig.~\ref{fig:results}(a)], the attempt loop is executed until a second photon is detected [point~$\mathit{III}$ in Fig.~\ref{fig:results}(a)].
After a variable delay $\Delta t$, both the memory and the network qubit are measured.
Note that no dynamical decoupling is used throughout this sequence.
Figure~\ref{fig:results}(c) shows the fidelity of ion-photon states to the closest maximally entangled state~\cite{badziag_local_2000} for different storage durations.
The raw \Sr{88}-photon fidelity is \ZeemanFidelity{}, but dephasing of the network qubit limits the coherence time of this state to \ZeemanLifetime{}.
Swapping the ion state into the memory qubit extends the coherence time by a factor \ish\num{70} with an initial fidelity of \ClockFidelity{}.
The additional infidelity is due to the high heating rates limiting the iSWAP operation~\cite{supplement}, and imperfections in the $\mathrm{\Stretch{}}\rightarrow \mathrm{\Clock{}}$ transfer pulse sequence~\cite{supplement}.
The fidelity shown in Fig.~\ref{fig:results}(c) decays due to magnetic field noise and laser leakage; heating during the storage duration causes single-qubit rotation errors in \Sr{88}, whereas the use of a co-propagating Raman beam geometry eliminates this effect in \Ca{43}.

In a second experiment, we demonstrate that these limitations can be overcome.
We employ Knill dynamical decoupling~\cite{souza_robust_2011, wang_single_2021} with \num{40} spin flips to suppress the effect of magnetic field noise [Fig.~\ref{fig:results}(b)].
To minimize the effect of laser leakage, we transport the ions \SI{100}{\micro\meter} away from the laser interaction zone.
Furthermore, sympathetic Doppler cooling on \Sr{88} avoids ion loss due to heating.
We achieve an \ac{IPE} fidelity of \TenSecFidelity{} after \SI{10}{\second} [squares and inset in Fig.~\ref{fig:results}(c)].
The ratio of decoherence rate to the node-to-node entanglement rate in a quantum network strongly impacts the resource scaling for fault-tolerant error correction~\cite{monroe_large_2014}.
Here, this ratio is estimated to be $0.0006$ and $0.08$ with and without dynamical decoupling, respectively, assuming the entanglement rate of \SI{182}{\per\second} previously observed in our setup~\cite{stephenson_high-rate_2020}. % includes factor 2 for decoherence of entangled 2-qubit state.

\begin{figure}[h!tb]
    \centering
    \includegraphics{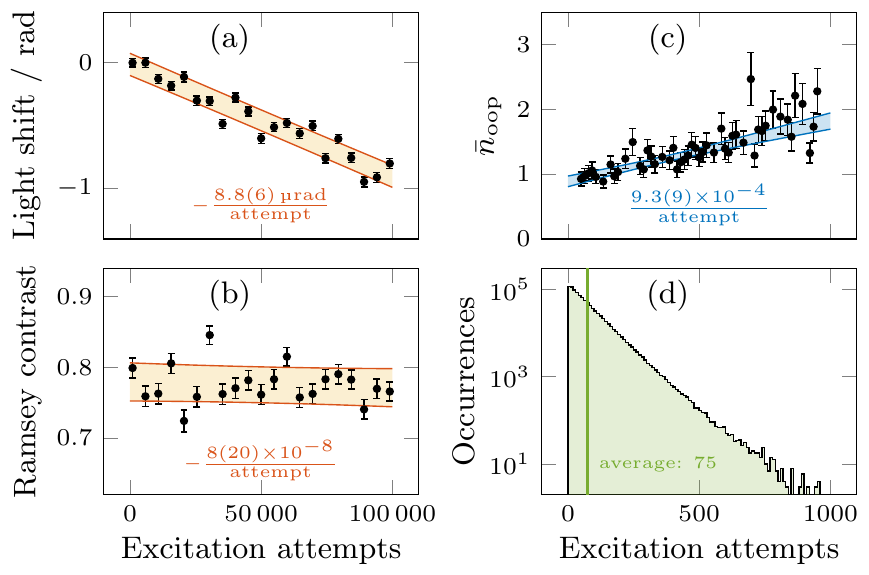}
    \caption{
    The effect of excitation attempts on (a) the phase and (b) coherence of the memory qubit, and on (c) the temperature of the \ac{OOP} mode, is probed by executing the attempt loop for a variable fraction of a fixed total duration.
    Each attempt takes \SI{1}{\micro\second}.
    The sensitivity to attempts, extracted from linear least-squares fits, is shown alongside the data.
    Error intervals span one standard deviation.
    (a-b) Ramsey experiments with \SI{100}{\milli\second} total duration.
    Filled single-prediction bands guide the eye to (a) phase and (b) contrast of the Ramsey fringe.
    (c) Sideband-ratio thermometry with \SI{1}{\milli\second} total duration.
    The \SI{95}{\percent} confidence band is shown in blue.
    (d) Histogram of excitation attempts until detection of a photon, indicating a success probability of $0.013$.
    }
    \label{fig:excitation}
\end{figure}

Crucially, there is negligible memory error associated with generating a second ion-photon pair, as the lasers used during the attempt loop are far off-resonant ($>\si{\tera\hertz}$) from transitions in \Ca{43}.
To demonstrate this, we perform Ramsey experiments on the memory qubit while the loop is ongoing in the background for up to $10^5$ excitation attempts [Fig.~\ref{fig:excitation}(a) and Fig.~\ref{fig:excitation}(b)], enough to herald $>\num{1500}$ entangled ion-photon states.
The light shift per excitation attempt is \LightShiftPerAttempt{}, and can easily be corrected in real-time by adjusting the phase reference.
From the same data, we do not observe any statistically significant reduction in contrast [Fig.~\ref{fig:excitation}(b)].
A secondary consequence of the loop is excess heating due to photon recoil.
We measure excess heating of \ExcessHeating{} phonons per attempt [Fig.~\ref{fig:excitation}(c)], which is insignificant in the context of this experiment [Fig.~\ref{fig:excitation}(d)].

In summary, we have demonstrated the coherent transfer of \ac{IPE} from a network qubit in \Sr{88} to a memory qubit in \Ca{43} within a quantum network node.
We note that the measurements reveal the presence of entanglement even though the photon was destroyed before the transfer took place~\cite{megidish_entanglement_2013}.
We extend the storage duration of this entanglement by \ish\num{4} orders of magnitude, to more than \SI{10}{\second}, while ensuring that subsequent \ac{IPE} can be performed without crosstalk affecting the memory qubit.
Extending the storage duration beyond the time taken to generate \ac{IPE} is essential for applications that require multiple communication photons.
We have shown that we can generate further \ac{IPE} on the network qubit while bounding the error induced on a memory qubit only \SI{3.3}{\micro\meter} away to $<\num{2e-5}$ [Fig.~\ref{fig:excitation}]; this enables applications such as entanglement distillation~\cite{bennett_purification_1996, monroe_scaling_2013, nigmatullin_minimally_2016} and blind quantum computing~\cite{broadbent_universal_2009}.
Mixed-species transfer in a network node also enables applications that require remote entanglement of long-lived memories, including quantum networks of atomic clocks~\cite{komar_quantum_2014, nichol_elementary_2022}.
For long-distance networks, communication latencies due to time-of-flight and classical signaling would limit the rate at which nodes with a single network qubit can generate entanglement.
However, if the state of this network qubit is stored in an available memory qubit immediately after emission of the photon, entanglement attempts could be made without dead-time in between~\cite{krutyanskiy_telecom-wavelength_2022}.
A constant attempt rate could be reached independent of distance, limited only by the local swapping procedure.
In that scheme, the memory qubits would be stored until the corresponding herald signals arrive to indicate which had been entangled successfully.
In our system, link losses, rather than memory coherence, would set the limit on the maximum possible node separation.
To increase the photon collection efficiency, cavities can be used~\cite{stute_quantum-state_2013, krutyanskiy_entanglement_2022}.
To reduce the fiber losses, quantum frequency conversion to infrared wavelengths has been proven feasible~\cite{wright_two-way_2018, krutyanskiy_light-matter_2019, hannegan_entanglement_2022}.
Combined with these improvements, our system, which integrates a high-fidelity photonic interface with mixed-species quantum logic, a robust memory and ion transport capabilities, paves the way for more powerful trapped-ion quantum networks.

% 7. Acknowledgements

We would like to thank Sandia National Laboratories for supplying HOA2 ion traps, the developers of the experiment control system ARTIQ~\cite{ARTIQ}, and C.~Matthiesen for technical assistance.
B.C.N.\ acknowledges funding from the U.K.\ National Physical Laboratory.
D.M.\ acknowledges support from St.~Anne's College Oxford.
G.A.M.\ acknowledges support from Wolfson College Oxford.
C.J.B.\ acknowledges support from a UKRI FL Fellowship, and is a Director of Oxford Ionics Ltd.
This work was supported by the U.K.\ EPSRC \enquote{Quantum Computing and Simulation} Hub, the E.U.\ Quantum Technology Flagship Project AQTION (No.\ 820495), and the U.S.\ Army Research Office (Ref.\ W911NF-18-1-0340).
{}

\begin{acronym}
    \acro{OOP}{out-of-phase}
    \acro{IP}{in-phase}
    \acro{IPE}{ion-photon entanglement}
\end{acronym}

\bibliography{library}

\begin{thebibliography}{62}%
\makeatletter
\providecommand \@ifxundefined [1]{%
 \@ifx{#1\undefined}
}%
\providecommand \@ifnum [1]{%
 \ifnum #1\expandafter \@firstoftwo
 \else \expandafter \@secondoftwo
 \fi
}%
\providecommand \@ifx [1]{%
 \ifx #1\expandafter \@firstoftwo
 \else \expandafter \@secondoftwo
 \fi
}%
\providecommand \natexlab [1]{#1}%
\providecommand \enquote  [1]{``#1''}%
\providecommand \bibnamefont  [1]{#1}%
\providecommand \bibfnamefont [1]{#1}%
\providecommand \citenamefont [1]{#1}%
\providecommand \href@noop [0]{\@secondoftwo}%
\providecommand \href [0]{\begingroup \@sanitize@url \@href}%
\providecommand \@href[1]{\@@startlink{#1}\@@href}%
\providecommand \@@href[1]{\endgroup#1\@@endlink}%
\providecommand \@sanitize@url [0]{\catcode `\\12\catcode `\$12\catcode
  `\&12\catcode `\#12\catcode `\^12\catcode `\_12\catcode `\%12\relax}%
\providecommand \@@startlink[1]{}%
\providecommand \@@endlink[0]{}%
\providecommand \url  [0]{\begingroup\@sanitize@url \@url }%
\providecommand \@url [1]{\endgroup\@href {#1}{\urlprefix }}%
\providecommand \urlprefix  [0]{URL }%
\providecommand \Eprint [0]{\href }%
\providecommand \doibase [0]{https://doi.org/}%
\providecommand \selectlanguage [0]{\@gobble}%
\providecommand \bibinfo  [0]{\@secondoftwo}%
\providecommand \bibfield  [0]{\@secondoftwo}%
\providecommand \translation [1]{[#1]}%
\providecommand \BibitemOpen [0]{}%
\providecommand \bibitemStop [0]{}%
\providecommand \bibitemNoStop [0]{.\EOS\space}%
\providecommand \EOS [0]{\spacefactor3000\relax}%
\providecommand \BibitemShut  [1]{\csname bibitem#1\endcsname}%
\let\auto@bib@innerbib\@empty
%</preamble>
\bibitem [{\citenamefont {Kimble}(2008)}]{kimble_quantum_2008}%
  \BibitemOpen
  \bibfield  {author} {\bibinfo {author} {\bibfnamefont {H.~J.}\ \bibnamefont
  {Kimble}},\ }\bibfield  {title} {\bibinfo {title} {The quantum internet},\
  }\href {http://www.nature.com/articles/nature07127} {\bibfield  {journal}
  {\bibinfo  {journal} {Nature}\ }\textbf {\bibinfo {volume} {453}},\ \bibinfo
  {pages} {1023} (\bibinfo {year} {2008})}\BibitemShut {NoStop}%
\bibitem [{\citenamefont {Bennett}\ and\ \citenamefont
  {Brassard}(1984)}]{bennett_quantum_1984}%
  \BibitemOpen
  \bibfield  {author} {\bibinfo {author} {\bibfnamefont {C.~H.}\ \bibnamefont
  {Bennett}}\ and\ \bibinfo {author} {\bibfnamefont {G.}~\bibnamefont
  {Brassard}},\ }\bibfield  {title} {\bibinfo {title} {Quantum cryptography:
  public key distribution and coin tossing},\ }\href@noop {} {\bibfield
  {journal} {\bibinfo  {journal} {Theor. Comput. Sci.}\ }\textbf {\bibinfo
  {volume} {560}},\ \bibinfo {pages} {7} (\bibinfo {year} {1984})}\BibitemShut
  {NoStop}%
\bibitem [{\citenamefont {Ekert}(1991)}]{ekert_quantum_1991}%
  \BibitemOpen
  \bibfield  {author} {\bibinfo {author} {\bibfnamefont {A.~K.}\ \bibnamefont
  {Ekert}},\ }\bibfield  {title} {\bibinfo {title} {Quantum cryptography based
  on bell's theorem},\ }\href {https://doi.org/10.1103/PhysRevLett.67.661}
  {\bibfield  {journal} {\bibinfo  {journal} {Phys. Rev. Lett.}\ }\textbf
  {\bibinfo {volume} {67}},\ \bibinfo {pages} {661} (\bibinfo {year}
  {1991})}\BibitemShut {NoStop}%
\bibitem [{\citenamefont {Bennett}\ \emph {et~al.}(1993)\citenamefont
  {Bennett}, \citenamefont {Brassard}, \citenamefont {Cr\'epeau}, \citenamefont
  {Jozsa}, \citenamefont {Peres},\ and\ \citenamefont
  {Wootters}}]{bennett_teleporting_1993}%
  \BibitemOpen
  \bibfield  {author} {\bibinfo {author} {\bibfnamefont {C.~H.}\ \bibnamefont
  {Bennett}}, \bibinfo {author} {\bibfnamefont {G.}~\bibnamefont {Brassard}},
  \bibinfo {author} {\bibfnamefont {C.}~\bibnamefont {Cr\'epeau}}, \bibinfo
  {author} {\bibfnamefont {R.}~\bibnamefont {Jozsa}}, \bibinfo {author}
  {\bibfnamefont {A.}~\bibnamefont {Peres}},\ and\ \bibinfo {author}
  {\bibfnamefont {W.~K.}\ \bibnamefont {Wootters}},\ }\bibfield  {title}
  {\bibinfo {title} {Teleporting an unknown quantum state via dual classical
  and einstein-podolsky-rosen channels},\ }\href
  {https://doi.org/10.1103/PhysRevLett.70.1895} {\bibfield  {journal} {\bibinfo
   {journal} {Phys. Rev. Lett.}\ }\textbf {\bibinfo {volume} {70}},\ \bibinfo
  {pages} {1895} (\bibinfo {year} {1993})}\BibitemShut {NoStop}%
\bibitem [{\citenamefont {Hucul}\ \emph {et~al.}(2015)\citenamefont {Hucul},
  \citenamefont {Inlek}, \citenamefont {Vittorini}, \citenamefont {Crocker},
  \citenamefont {Debnath}, \citenamefont {Clark},\ and\ \citenamefont
  {Monroe}}]{hucul_modular_2015}%
  \BibitemOpen
  \bibfield  {author} {\bibinfo {author} {\bibfnamefont {D.}~\bibnamefont
  {Hucul}}, \bibinfo {author} {\bibfnamefont {I.~V.}\ \bibnamefont {Inlek}},
  \bibinfo {author} {\bibfnamefont {G.}~\bibnamefont {Vittorini}}, \bibinfo
  {author} {\bibfnamefont {C.}~\bibnamefont {Crocker}}, \bibinfo {author}
  {\bibfnamefont {S.}~\bibnamefont {Debnath}}, \bibinfo {author} {\bibfnamefont
  {S.~M.}\ \bibnamefont {Clark}},\ and\ \bibinfo {author} {\bibfnamefont
  {C.}~\bibnamefont {Monroe}},\ }\bibfield  {title} {\bibinfo {title} {Modular
  entanglement of atomic qubits using photons and phonons},\ }\href
  {https://doi.org/10.1038/nphys3150} {\bibfield  {journal} {\bibinfo
  {journal} {Nat. Phys.}\ }\textbf {\bibinfo {volume} {11}},\ \bibinfo {pages}
  {37} (\bibinfo {year} {2015})}\BibitemShut {NoStop}%
\bibitem [{\citenamefont {K{\'{o}}m{\'{a}}r}\ \emph {et~al.}(2014)\citenamefont
  {K{\'{o}}m{\'{a}}r}, \citenamefont {Kessler}, \citenamefont {Bishof},
  \citenamefont {Jiang}, \citenamefont {S{\o}rensen}, \citenamefont {Ye},\ and\
  \citenamefont {Lukin}}]{komar_quantum_2014}%
  \BibitemOpen
  \bibfield  {author} {\bibinfo {author} {\bibfnamefont {P.}~\bibnamefont
  {K{\'{o}}m{\'{a}}r}}, \bibinfo {author} {\bibfnamefont {E.~M.}\ \bibnamefont
  {Kessler}}, \bibinfo {author} {\bibfnamefont {M.}~\bibnamefont {Bishof}},
  \bibinfo {author} {\bibfnamefont {L.}~\bibnamefont {Jiang}}, \bibinfo
  {author} {\bibfnamefont {A.~S.}\ \bibnamefont {S{\o}rensen}}, \bibinfo
  {author} {\bibfnamefont {J.}~\bibnamefont {Ye}},\ and\ \bibinfo {author}
  {\bibfnamefont {M.~D.}\ \bibnamefont {Lukin}},\ }\bibfield  {title} {\bibinfo
  {title} {A quantum network of clocks},\ }\href
  {http://www.nature.com/articles/nphys3000} {\bibfield  {journal} {\bibinfo
  {journal} {Nat. Phys.}\ }\textbf {\bibinfo {volume} {10}},\ \bibinfo {pages}
  {582} (\bibinfo {year} {2014})}\BibitemShut {NoStop}%
\bibitem [{\citenamefont {Matsukevich}\ \emph {et~al.}(2008)\citenamefont
  {Matsukevich}, \citenamefont {Maunz}, \citenamefont {Moehring}, \citenamefont
  {Olmschenk},\ and\ \citenamefont {Monroe}}]{matsukevich_bell_2008}%
  \BibitemOpen
  \bibfield  {author} {\bibinfo {author} {\bibfnamefont {D.~N.}\ \bibnamefont
  {Matsukevich}}, \bibinfo {author} {\bibfnamefont {P.}~\bibnamefont {Maunz}},
  \bibinfo {author} {\bibfnamefont {D.~L.}\ \bibnamefont {Moehring}}, \bibinfo
  {author} {\bibfnamefont {S.}~\bibnamefont {Olmschenk}},\ and\ \bibinfo
  {author} {\bibfnamefont {C.}~\bibnamefont {Monroe}},\ }\bibfield  {title}
  {\bibinfo {title} {Bell inequality violation with two remote atomic qubits},\
  }\href {https://doi.org/10.1103/PhysRevLett.100.150404} {\bibfield  {journal}
  {\bibinfo  {journal} {Phys. Rev. Lett.}\ }\textbf {\bibinfo {volume} {100}},\
  \bibinfo {pages} {150404} (\bibinfo {year} {2008})}\BibitemShut {NoStop}%
\bibitem [{\citenamefont {Hensen}\ \emph {et~al.}(2015)\citenamefont {Hensen},
  \citenamefont {Bernien}, \citenamefont {Dr{\'{e}}au}, \citenamefont
  {Reiserer}, \citenamefont {Kalb}, \citenamefont {Blok}, \citenamefont
  {Ruitenberg}, \citenamefont {Vermeulen}, \citenamefont {Schouten},
  \citenamefont {Abell{\'{a}}n}, \citenamefont {Amaya}, \citenamefont
  {Pruneri}, \citenamefont {Mitchell}, \citenamefont {Markham}, \citenamefont
  {Twitchen}, \citenamefont {Elkouss}, \citenamefont {Wehner}, \citenamefont
  {Taminiau},\ and\ \citenamefont {Hanson}}]{hensen_loophole-free_2015}%
  \BibitemOpen
  \bibfield  {author} {\bibinfo {author} {\bibfnamefont {B.}~\bibnamefont
  {Hensen}}, \bibinfo {author} {\bibfnamefont {H.}~\bibnamefont {Bernien}},
  \bibinfo {author} {\bibfnamefont {A.~E.}\ \bibnamefont {Dr{\'{e}}au}},
  \bibinfo {author} {\bibfnamefont {A.}~\bibnamefont {Reiserer}}, \bibinfo
  {author} {\bibfnamefont {N.}~\bibnamefont {Kalb}}, \bibinfo {author}
  {\bibfnamefont {M.~S.}\ \bibnamefont {Blok}}, \bibinfo {author}
  {\bibfnamefont {J.}~\bibnamefont {Ruitenberg}}, \bibinfo {author}
  {\bibfnamefont {R.~F.~L.}\ \bibnamefont {Vermeulen}}, \bibinfo {author}
  {\bibfnamefont {R.~N.}\ \bibnamefont {Schouten}}, \bibinfo {author}
  {\bibfnamefont {C.}~\bibnamefont {Abell{\'{a}}n}}, \bibinfo {author}
  {\bibfnamefont {W.}~\bibnamefont {Amaya}}, \bibinfo {author} {\bibfnamefont
  {V.}~\bibnamefont {Pruneri}}, \bibinfo {author} {\bibfnamefont {M.~W.}\
  \bibnamefont {Mitchell}}, \bibinfo {author} {\bibfnamefont {M.}~\bibnamefont
  {Markham}}, \bibinfo {author} {\bibfnamefont {D.~J.}\ \bibnamefont
  {Twitchen}}, \bibinfo {author} {\bibfnamefont {D.}~\bibnamefont {Elkouss}},
  \bibinfo {author} {\bibfnamefont {S.}~\bibnamefont {Wehner}}, \bibinfo
  {author} {\bibfnamefont {T.~H.}\ \bibnamefont {Taminiau}},\ and\ \bibinfo
  {author} {\bibfnamefont {R.}~\bibnamefont {Hanson}},\ }\bibfield  {title}
  {\bibinfo {title} {Loophole-free {Bell} inequality violation using electron
  spins separated by 1.3 kilometres},\ }\href
  {http://www.nature.com/articles/nature15759} {\bibfield  {journal} {\bibinfo
  {journal} {Nature}\ }\textbf {\bibinfo {volume} {526}},\ \bibinfo {pages}
  {682} (\bibinfo {year} {2015})}\BibitemShut {NoStop}%
\bibitem [{\citenamefont {Northup}\ and\ \citenamefont
  {Blatt}(2014)}]{northup_quantum_2014}%
  \BibitemOpen
  \bibfield  {author} {\bibinfo {author} {\bibfnamefont {T.~E.}\ \bibnamefont
  {Northup}}\ and\ \bibinfo {author} {\bibfnamefont {R.}~\bibnamefont
  {Blatt}},\ }\bibfield  {title} {\bibinfo {title} {Quantum information
  transfer using photons},\ }\href {https://doi.org/10.1038/nphoton.2014.53}
  {\bibfield  {journal} {\bibinfo  {journal} {Nat. Photon.}\ }\textbf {\bibinfo
  {volume} {8}},\ \bibinfo {pages} {356} (\bibinfo {year} {2014})}\BibitemShut
  {NoStop}%
\bibitem [{\citenamefont {Hermans}\ \emph {et~al.}(2022)\citenamefont
  {Hermans}, \citenamefont {Pompili}, \citenamefont {Beukers}, \citenamefont
  {Baier}, \citenamefont {Borregaard},\ and\ \citenamefont
  {Hanson}}]{hermans_qubit_2022}%
  \BibitemOpen
  \bibfield  {author} {\bibinfo {author} {\bibfnamefont {S.~L.~N.}\
  \bibnamefont {Hermans}}, \bibinfo {author} {\bibfnamefont {M.}~\bibnamefont
  {Pompili}}, \bibinfo {author} {\bibfnamefont {H.~K.~C.}\ \bibnamefont
  {Beukers}}, \bibinfo {author} {\bibfnamefont {S.}~\bibnamefont {Baier}},
  \bibinfo {author} {\bibfnamefont {J.}~\bibnamefont {Borregaard}},\ and\
  \bibinfo {author} {\bibfnamefont {R.}~\bibnamefont {Hanson}},\ }\bibfield
  {title} {\bibinfo {title} {Qubit teleportation between non-neighbouring nodes
  in a quantum network},\ }\href
  {https://www.nature.com/articles/s41586-022-04697-y} {\bibfield  {journal}
  {\bibinfo  {journal} {Nature}\ }\textbf {\bibinfo {volume} {605}},\ \bibinfo
  {pages} {663} (\bibinfo {year} {2022})}\BibitemShut {NoStop}%
\bibitem [{\citenamefont {Jennewein}\ \emph {et~al.}(2000)\citenamefont
  {Jennewein}, \citenamefont {Simon}, \citenamefont {Weihs}, \citenamefont
  {Weinfurter},\ and\ \citenamefont {Zeilinger}}]{jennewein_quantum_2000}%
  \BibitemOpen
  \bibfield  {author} {\bibinfo {author} {\bibfnamefont {T.}~\bibnamefont
  {Jennewein}}, \bibinfo {author} {\bibfnamefont {C.}~\bibnamefont {Simon}},
  \bibinfo {author} {\bibfnamefont {G.}~\bibnamefont {Weihs}}, \bibinfo
  {author} {\bibfnamefont {H.}~\bibnamefont {Weinfurter}},\ and\ \bibinfo
  {author} {\bibfnamefont {A.}~\bibnamefont {Zeilinger}},\ }\bibfield  {title}
  {\bibinfo {title} {Quantum cryptography with entangled photons},\ }\href
  {https://doi.org/10.1103/PhysRevLett.84.4729} {\bibfield  {journal} {\bibinfo
   {journal} {Phys. Rev. Lett.}\ }\textbf {\bibinfo {volume} {84}},\ \bibinfo
  {pages} {4729} (\bibinfo {year} {2000})}\BibitemShut {NoStop}%
\bibitem [{\citenamefont {Yin}\ \emph {et~al.}(2012)\citenamefont {Yin},
  \citenamefont {Ren}, \citenamefont {Lu}, \citenamefont {Cao}, \citenamefont
  {Yong}, \citenamefont {Wu}, \citenamefont {Liu}, \citenamefont {Liao},
  \citenamefont {Zhou}, \citenamefont {Jiang}, \citenamefont {Cai},
  \citenamefont {Xu}, \citenamefont {Pan}, \citenamefont {Jia}, \citenamefont
  {Huang}, \citenamefont {Yin}, \citenamefont {Wang}, \citenamefont {Chen},
  \citenamefont {Peng},\ and\ \citenamefont {Pan}}]{yin_quantum_2012}%
  \BibitemOpen
  \bibfield  {author} {\bibinfo {author} {\bibfnamefont {J.}~\bibnamefont
  {Yin}}, \bibinfo {author} {\bibfnamefont {J.-G.}\ \bibnamefont {Ren}},
  \bibinfo {author} {\bibfnamefont {H.}~\bibnamefont {Lu}}, \bibinfo {author}
  {\bibfnamefont {Y.}~\bibnamefont {Cao}}, \bibinfo {author} {\bibfnamefont
  {H.-L.}\ \bibnamefont {Yong}}, \bibinfo {author} {\bibfnamefont {Y.-P.}\
  \bibnamefont {Wu}}, \bibinfo {author} {\bibfnamefont {C.}~\bibnamefont
  {Liu}}, \bibinfo {author} {\bibfnamefont {S.-K.}\ \bibnamefont {Liao}},
  \bibinfo {author} {\bibfnamefont {F.}~\bibnamefont {Zhou}}, \bibinfo {author}
  {\bibfnamefont {Y.}~\bibnamefont {Jiang}}, \bibinfo {author} {\bibfnamefont
  {X.-D.}\ \bibnamefont {Cai}}, \bibinfo {author} {\bibfnamefont
  {P.}~\bibnamefont {Xu}}, \bibinfo {author} {\bibfnamefont {G.-S.}\
  \bibnamefont {Pan}}, \bibinfo {author} {\bibfnamefont {J.-J.}\ \bibnamefont
  {Jia}}, \bibinfo {author} {\bibfnamefont {Y.-M.}\ \bibnamefont {Huang}},
  \bibinfo {author} {\bibfnamefont {H.}~\bibnamefont {Yin}}, \bibinfo {author}
  {\bibfnamefont {J.-Y.}\ \bibnamefont {Wang}}, \bibinfo {author}
  {\bibfnamefont {Y.-A.}\ \bibnamefont {Chen}}, \bibinfo {author}
  {\bibfnamefont {C.-Z.}\ \bibnamefont {Peng}},\ and\ \bibinfo {author}
  {\bibfnamefont {J.-W.}\ \bibnamefont {Pan}},\ }\bibfield  {title} {\bibinfo
  {title} {Quantum teleportation and entanglement distribution over
  100-kilometre free-space channels},\ }\href
  {http://www.nature.com/articles/nature11332} {\bibfield  {journal} {\bibinfo
  {journal} {Nature}\ }\textbf {\bibinfo {volume} {488}},\ \bibinfo {pages}
  {185} (\bibinfo {year} {2012})}\BibitemShut {NoStop}%
\bibitem [{\citenamefont {Matsukevich}\ \emph {et~al.}(2006)\citenamefont
  {Matsukevich}, \citenamefont {Chaneli\`ere}, \citenamefont {Jenkins},
  \citenamefont {Lan}, \citenamefont {Kennedy},\ and\ \citenamefont
  {Kuzmich}}]{matsukevich_entanglement_2006}%
  \BibitemOpen
  \bibfield  {author} {\bibinfo {author} {\bibfnamefont {D.~N.}\ \bibnamefont
  {Matsukevich}}, \bibinfo {author} {\bibfnamefont {T.}~\bibnamefont
  {Chaneli\`ere}}, \bibinfo {author} {\bibfnamefont {S.~D.}\ \bibnamefont
  {Jenkins}}, \bibinfo {author} {\bibfnamefont {S.-Y.}\ \bibnamefont {Lan}},
  \bibinfo {author} {\bibfnamefont {T.~A.~B.}\ \bibnamefont {Kennedy}},\ and\
  \bibinfo {author} {\bibfnamefont {A.}~\bibnamefont {Kuzmich}},\ }\bibfield
  {title} {\bibinfo {title} {Entanglement of remote atomic qubits},\ }\href
  {https://doi.org/10.1103/PhysRevLett.96.030405} {\bibfield  {journal}
  {\bibinfo  {journal} {Phys. Rev. Lett.}\ }\textbf {\bibinfo {volume} {96}},\
  \bibinfo {pages} {030405} (\bibinfo {year} {2006})}\BibitemShut {NoStop}%
\bibitem [{\citenamefont {N\"olleke}\ \emph {et~al.}(2013)\citenamefont
  {N\"olleke}, \citenamefont {Neuzner}, \citenamefont {Reiserer}, \citenamefont
  {Hahn}, \citenamefont {Rempe},\ and\ \citenamefont
  {Ritter}}]{nolleke_efficient_2013}%
  \BibitemOpen
  \bibfield  {author} {\bibinfo {author} {\bibfnamefont {C.}~\bibnamefont
  {N\"olleke}}, \bibinfo {author} {\bibfnamefont {A.}~\bibnamefont {Neuzner}},
  \bibinfo {author} {\bibfnamefont {A.}~\bibnamefont {Reiserer}}, \bibinfo
  {author} {\bibfnamefont {C.}~\bibnamefont {Hahn}}, \bibinfo {author}
  {\bibfnamefont {G.}~\bibnamefont {Rempe}},\ and\ \bibinfo {author}
  {\bibfnamefont {S.}~\bibnamefont {Ritter}},\ }\bibfield  {title} {\bibinfo
  {title} {Efficient teleportation between remote single-atom quantum
  memories},\ }\href {https://doi.org/10.1103/PhysRevLett.110.140403}
  {\bibfield  {journal} {\bibinfo  {journal} {Phys. Rev. Lett.}\ }\textbf
  {\bibinfo {volume} {110}},\ \bibinfo {pages} {140403} (\bibinfo {year}
  {2013})}\BibitemShut {NoStop}%
\bibitem [{\citenamefont {Zhang}\ \emph {et~al.}(2022)\citenamefont {Zhang},
  \citenamefont {van Leent}, \citenamefont {Redeker}, \citenamefont {Garthoff},
  \citenamefont {Schwonnek}, \citenamefont {Fertig}, \citenamefont {Eppelt},
  \citenamefont {Rosenfeld}, \citenamefont {Scarani}, \citenamefont {Lim},\
  and\ \citenamefont {Weinfurter}}]{zhang_device-independent_2022}%
  \BibitemOpen
  \bibfield  {author} {\bibinfo {author} {\bibfnamefont {W.}~\bibnamefont
  {Zhang}}, \bibinfo {author} {\bibfnamefont {T.}~\bibnamefont {van Leent}},
  \bibinfo {author} {\bibfnamefont {K.}~\bibnamefont {Redeker}}, \bibinfo
  {author} {\bibfnamefont {R.}~\bibnamefont {Garthoff}}, \bibinfo {author}
  {\bibfnamefont {R.}~\bibnamefont {Schwonnek}}, \bibinfo {author}
  {\bibfnamefont {F.}~\bibnamefont {Fertig}}, \bibinfo {author} {\bibfnamefont
  {S.}~\bibnamefont {Eppelt}}, \bibinfo {author} {\bibfnamefont
  {W.}~\bibnamefont {Rosenfeld}}, \bibinfo {author} {\bibfnamefont
  {V.}~\bibnamefont {Scarani}}, \bibinfo {author} {\bibfnamefont {C.~C.-W.}\
  \bibnamefont {Lim}},\ and\ \bibinfo {author} {\bibfnamefont {H.}~\bibnamefont
  {Weinfurter}},\ }\bibfield  {title} {\bibinfo {title} {A device-independent
  quantum key distribution system for distant users},\ }\href
  {https://www.nature.com/articles/s41586-022-04891-y} {\bibfield  {journal}
  {\bibinfo  {journal} {Nature}\ }\textbf {\bibinfo {volume} {607}},\ \bibinfo
  {pages} {687} (\bibinfo {year} {2022})}\BibitemShut {NoStop}%
\bibitem [{\citenamefont {Magnard}\ \emph {et~al.}(2020)\citenamefont
  {Magnard}, \citenamefont {Storz}, \citenamefont {Kurpiers}, \citenamefont
  {Sch\"ar}, \citenamefont {Marxer}, \citenamefont {L\"utolf}, \citenamefont
  {Walter}, \citenamefont {Besse}, \citenamefont {Gabureac}, \citenamefont
  {Reuer}, \citenamefont {Akin}, \citenamefont {Royer}, \citenamefont {Blais},\
  and\ \citenamefont {Wallraff}}]{magnard_microwave_2020}%
  \BibitemOpen
  \bibfield  {author} {\bibinfo {author} {\bibfnamefont {P.}~\bibnamefont
  {Magnard}}, \bibinfo {author} {\bibfnamefont {S.}~\bibnamefont {Storz}},
  \bibinfo {author} {\bibfnamefont {P.}~\bibnamefont {Kurpiers}}, \bibinfo
  {author} {\bibfnamefont {J.}~\bibnamefont {Sch\"ar}}, \bibinfo {author}
  {\bibfnamefont {F.}~\bibnamefont {Marxer}}, \bibinfo {author} {\bibfnamefont
  {J.}~\bibnamefont {L\"utolf}}, \bibinfo {author} {\bibfnamefont
  {T.}~\bibnamefont {Walter}}, \bibinfo {author} {\bibfnamefont {J.-C.}\
  \bibnamefont {Besse}}, \bibinfo {author} {\bibfnamefont {M.}~\bibnamefont
  {Gabureac}}, \bibinfo {author} {\bibfnamefont {K.}~\bibnamefont {Reuer}},
  \bibinfo {author} {\bibfnamefont {A.}~\bibnamefont {Akin}}, \bibinfo {author}
  {\bibfnamefont {B.}~\bibnamefont {Royer}}, \bibinfo {author} {\bibfnamefont
  {A.}~\bibnamefont {Blais}},\ and\ \bibinfo {author} {\bibfnamefont
  {A.}~\bibnamefont {Wallraff}},\ }\bibfield  {title} {\bibinfo {title}
  {Microwave quantum link between superconducting circuits housed in spatially
  separated cryogenic systems},\ }\href
  {https://doi.org/10.1103/PhysRevLett.125.260502} {\bibfield  {journal}
  {\bibinfo  {journal} {Phys. Rev. Lett.}\ }\textbf {\bibinfo {volume} {125}},\
  \bibinfo {pages} {260502} (\bibinfo {year} {2020})}\BibitemShut {NoStop}%
\bibitem [{\citenamefont {Moehring}\ \emph {et~al.}(2007)\citenamefont
  {Moehring}, \citenamefont {Maunz}, \citenamefont {Olmschenk}, \citenamefont
  {Younge}, \citenamefont {Matsukevich}, \citenamefont {Duan},\ and\
  \citenamefont {Monroe}}]{moehring_entanglement_2007}%
  \BibitemOpen
  \bibfield  {author} {\bibinfo {author} {\bibfnamefont {D.~L.}\ \bibnamefont
  {Moehring}}, \bibinfo {author} {\bibfnamefont {P.}~\bibnamefont {Maunz}},
  \bibinfo {author} {\bibfnamefont {S.}~\bibnamefont {Olmschenk}}, \bibinfo
  {author} {\bibfnamefont {K.~C.}\ \bibnamefont {Younge}}, \bibinfo {author}
  {\bibfnamefont {D.~N.}\ \bibnamefont {Matsukevich}}, \bibinfo {author}
  {\bibfnamefont {L.-M.}\ \bibnamefont {Duan}},\ and\ \bibinfo {author}
  {\bibfnamefont {C.}~\bibnamefont {Monroe}},\ }\bibfield  {title} {\bibinfo
  {title} {Entanglement of single-atom quantum bits at a distance},\ }\href
  {https://www.nature.com/articles/nature06118} {\bibfield  {journal} {\bibinfo
   {journal} {Nature}\ }\textbf {\bibinfo {volume} {449}},\ \bibinfo {pages}
  {68} (\bibinfo {year} {2007})}\BibitemShut {NoStop}%
\bibitem [{\citenamefont {Olmschenk}\ \emph {et~al.}(2009)\citenamefont
  {Olmschenk}, \citenamefont {Matsukevich}, \citenamefont {Maunz},
  \citenamefont {Hayes}, \citenamefont {Duan},\ and\ \citenamefont
  {Monroe}}]{olmschenk_quantum_2009}%
  \BibitemOpen
  \bibfield  {author} {\bibinfo {author} {\bibfnamefont {S.}~\bibnamefont
  {Olmschenk}}, \bibinfo {author} {\bibfnamefont {D.~N.}\ \bibnamefont
  {Matsukevich}}, \bibinfo {author} {\bibfnamefont {P.}~\bibnamefont {Maunz}},
  \bibinfo {author} {\bibfnamefont {D.}~\bibnamefont {Hayes}}, \bibinfo
  {author} {\bibfnamefont {L.-M.}\ \bibnamefont {Duan}},\ and\ \bibinfo
  {author} {\bibfnamefont {C.}~\bibnamefont {Monroe}},\ }\bibfield  {title}
  {\bibinfo {title} {Quantum teleportation between distant matter qubits},\
  }\href {https://doi.org/10.1126/science.1167209} {\bibfield  {journal}
  {\bibinfo  {journal} {Science}\ }\textbf {\bibinfo {volume} {323}},\ \bibinfo
  {pages} {486} (\bibinfo {year} {2009})}\BibitemShut {NoStop}%
\bibitem [{\citenamefont {Pironio}\ \emph {et~al.}(2010)\citenamefont
  {Pironio}, \citenamefont {Ac{\'{i}}n}, \citenamefont {Massar}, \citenamefont
  {de~la Giroday}, \citenamefont {Matsukevich}, \citenamefont {Maunz},
  \citenamefont {Olmschenk}, \citenamefont {Hayes}, \citenamefont {Luo},
  \citenamefont {Manning},\ and\ \citenamefont {Monroe}}]{pironio_random_2010}%
  \BibitemOpen
  \bibfield  {author} {\bibinfo {author} {\bibfnamefont {S.}~\bibnamefont
  {Pironio}}, \bibinfo {author} {\bibfnamefont {A.}~\bibnamefont {Ac{\'{i}}n}},
  \bibinfo {author} {\bibfnamefont {S.}~\bibnamefont {Massar}}, \bibinfo
  {author} {\bibfnamefont {A.~B.}\ \bibnamefont {de~la Giroday}}, \bibinfo
  {author} {\bibfnamefont {D.~N.}\ \bibnamefont {Matsukevich}}, \bibinfo
  {author} {\bibfnamefont {P.}~\bibnamefont {Maunz}}, \bibinfo {author}
  {\bibfnamefont {S.}~\bibnamefont {Olmschenk}}, \bibinfo {author}
  {\bibfnamefont {D.}~\bibnamefont {Hayes}}, \bibinfo {author} {\bibfnamefont
  {L.}~\bibnamefont {Luo}}, \bibinfo {author} {\bibfnamefont {T.~A.}\
  \bibnamefont {Manning}},\ and\ \bibinfo {author} {\bibfnamefont
  {C.}~\bibnamefont {Monroe}},\ }\bibfield  {title} {\bibinfo {title} {Random
  numbers certified by {Bell}’s theorem},\ }\href
  {http://www.nature.com/articles/nature09008} {\bibfield  {journal} {\bibinfo
  {journal} {Nature}\ }\textbf {\bibinfo {volume} {464}},\ \bibinfo {pages}
  {1021} (\bibinfo {year} {2010})}\BibitemShut {NoStop}%
\bibitem [{\citenamefont {Stephenson}\ \emph {et~al.}(2020)\citenamefont
  {Stephenson}, \citenamefont {Nadlinger}, \citenamefont {Nichol},
  \citenamefont {An}, \citenamefont {Drmota}, \citenamefont {Ballance},
  \citenamefont {Thirumalai}, \citenamefont {Goodwin}, \citenamefont {Lucas},\
  and\ \citenamefont {Ballance}}]{stephenson_high-rate_2020}%
  \BibitemOpen
  \bibfield  {author} {\bibinfo {author} {\bibfnamefont {L.~J.}\ \bibnamefont
  {Stephenson}}, \bibinfo {author} {\bibfnamefont {D.~P.}\ \bibnamefont
  {Nadlinger}}, \bibinfo {author} {\bibfnamefont {B.~C.}\ \bibnamefont
  {Nichol}}, \bibinfo {author} {\bibfnamefont {S.}~\bibnamefont {An}}, \bibinfo
  {author} {\bibfnamefont {P.}~\bibnamefont {Drmota}}, \bibinfo {author}
  {\bibfnamefont {T.~G.}\ \bibnamefont {Ballance}}, \bibinfo {author}
  {\bibfnamefont {K.}~\bibnamefont {Thirumalai}}, \bibinfo {author}
  {\bibfnamefont {J.~F.}\ \bibnamefont {Goodwin}}, \bibinfo {author}
  {\bibfnamefont {D.~M.}\ \bibnamefont {Lucas}},\ and\ \bibinfo {author}
  {\bibfnamefont {C.~J.}\ \bibnamefont {Ballance}},\ }\bibfield  {title}
  {\bibinfo {title} {High-rate, high-fidelity entanglement of qubits across an
  elementary quantum network},\ }\href
  {https://doi.org/10.1103/PhysRevLett.124.110501} {\bibfield  {journal}
  {\bibinfo  {journal} {Phys. Rev. Lett.}\ }\textbf {\bibinfo {volume} {124}},\
  \bibinfo {pages} {110501} (\bibinfo {year} {2020})}\BibitemShut {NoStop}%
\bibitem [{\citenamefont {Nadlinger}\ \emph {et~al.}(2022)\citenamefont
  {Nadlinger}, \citenamefont {Drmota}, \citenamefont {Nichol}, \citenamefont
  {Araneda}, \citenamefont {Main}, \citenamefont {Srinivas}, \citenamefont
  {Lucas}, \citenamefont {Ballance}, \citenamefont {Ivanov}, \citenamefont
  {Tan}, \citenamefont {Sekatski}, \citenamefont {Urbanke}, \citenamefont
  {Renner}, \citenamefont {Sangouard},\ and\ \citenamefont
  {Bancal}}]{nadlinger_experimental_2022}%
  \BibitemOpen
  \bibfield  {author} {\bibinfo {author} {\bibfnamefont {D.~P.}\ \bibnamefont
  {Nadlinger}}, \bibinfo {author} {\bibfnamefont {P.}~\bibnamefont {Drmota}},
  \bibinfo {author} {\bibfnamefont {B.~C.}\ \bibnamefont {Nichol}}, \bibinfo
  {author} {\bibfnamefont {G.}~\bibnamefont {Araneda}}, \bibinfo {author}
  {\bibfnamefont {D.}~\bibnamefont {Main}}, \bibinfo {author} {\bibfnamefont
  {R.}~\bibnamefont {Srinivas}}, \bibinfo {author} {\bibfnamefont {D.~M.}\
  \bibnamefont {Lucas}}, \bibinfo {author} {\bibfnamefont {C.~J.}\ \bibnamefont
  {Ballance}}, \bibinfo {author} {\bibfnamefont {K.}~\bibnamefont {Ivanov}},
  \bibinfo {author} {\bibfnamefont {E.~Y.-Z.}\ \bibnamefont {Tan}}, \bibinfo
  {author} {\bibfnamefont {P.}~\bibnamefont {Sekatski}}, \bibinfo {author}
  {\bibfnamefont {R.~L.}\ \bibnamefont {Urbanke}}, \bibinfo {author}
  {\bibfnamefont {R.}~\bibnamefont {Renner}}, \bibinfo {author} {\bibfnamefont
  {N.}~\bibnamefont {Sangouard}},\ and\ \bibinfo {author} {\bibfnamefont
  {J.-D.}\ \bibnamefont {Bancal}},\ }\bibfield  {title} {\bibinfo {title}
  {Experimental quantum key distribution certified by {Bell}'s theorem},\
  }\href {https://www.nature.com/articles/s41586-022-04941-5} {\bibfield
  {journal} {\bibinfo  {journal} {Nature}\ }\textbf {\bibinfo {volume} {607}},\
  \bibinfo {pages} {682} (\bibinfo {year} {2022})}\BibitemShut {NoStop}%
\bibitem [{\citenamefont {Nichol}\ \emph {et~al.}(2022)\citenamefont {Nichol},
  \citenamefont {Srinivas}, \citenamefont {Nadlinger}, \citenamefont {Drmota},
  \citenamefont {Main}, \citenamefont {Araneda}, \citenamefont {Ballance},\
  and\ \citenamefont {Lucas}}]{nichol_elementary_2022}%
  \BibitemOpen
  \bibfield  {author} {\bibinfo {author} {\bibfnamefont {B.~C.}\ \bibnamefont
  {Nichol}}, \bibinfo {author} {\bibfnamefont {R.}~\bibnamefont {Srinivas}},
  \bibinfo {author} {\bibfnamefont {D.~P.}\ \bibnamefont {Nadlinger}}, \bibinfo
  {author} {\bibfnamefont {P.}~\bibnamefont {Drmota}}, \bibinfo {author}
  {\bibfnamefont {D.}~\bibnamefont {Main}}, \bibinfo {author} {\bibfnamefont
  {G.}~\bibnamefont {Araneda}}, \bibinfo {author} {\bibfnamefont {C.~J.}\
  \bibnamefont {Ballance}},\ and\ \bibinfo {author} {\bibfnamefont {D.~M.}\
  \bibnamefont {Lucas}},\ }\bibfield  {title} {\bibinfo {title} {An elementary
  quantum network of entangled optical atomic clocks},\ }\href
  {https://www.nature.com/articles/s41586-022-05088-z} {\bibfield  {journal}
  {\bibinfo  {journal} {Nature}\ }\textbf {\bibinfo {volume} {609}},\ \bibinfo
  {pages} {689} (\bibinfo {year} {2022})}\BibitemShut {NoStop}%
\bibitem [{\citenamefont {Krutyanskiy}\ \emph {et~al.}(2023)\citenamefont
  {Krutyanskiy}, \citenamefont {Galli}, \citenamefont {Krcmarsky},
  \citenamefont {Baier}, \citenamefont {Fioretto}, \citenamefont {Pu},
  \citenamefont {Mazloom}, \citenamefont {Sekatski}, \citenamefont {Canteri},
  \citenamefont {Teller}, \citenamefont {Schupp}, \citenamefont {Bate},
  \citenamefont {Meraner}, \citenamefont {Sangouard}, \citenamefont {Lanyon},\
  and\ \citenamefont {Northup}}]{krutyanskiy_entanglement_2022}%
  \BibitemOpen
  \bibfield  {author} {\bibinfo {author} {\bibfnamefont {V.}~\bibnamefont
  {Krutyanskiy}}, \bibinfo {author} {\bibfnamefont {M.}~\bibnamefont {Galli}},
  \bibinfo {author} {\bibfnamefont {V.}~\bibnamefont {Krcmarsky}}, \bibinfo
  {author} {\bibfnamefont {S.}~\bibnamefont {Baier}}, \bibinfo {author}
  {\bibfnamefont {D.~A.}\ \bibnamefont {Fioretto}}, \bibinfo {author}
  {\bibfnamefont {Y.}~\bibnamefont {Pu}}, \bibinfo {author} {\bibfnamefont
  {A.}~\bibnamefont {Mazloom}}, \bibinfo {author} {\bibfnamefont
  {P.}~\bibnamefont {Sekatski}}, \bibinfo {author} {\bibfnamefont
  {M.}~\bibnamefont {Canteri}}, \bibinfo {author} {\bibfnamefont
  {M.}~\bibnamefont {Teller}}, \bibinfo {author} {\bibfnamefont
  {J.}~\bibnamefont {Schupp}}, \bibinfo {author} {\bibfnamefont
  {J.}~\bibnamefont {Bate}}, \bibinfo {author} {\bibfnamefont {M.}~\bibnamefont
  {Meraner}}, \bibinfo {author} {\bibfnamefont {N.}~\bibnamefont {Sangouard}},
  \bibinfo {author} {\bibfnamefont {B.~P.}\ \bibnamefont {Lanyon}},\ and\
  \bibinfo {author} {\bibfnamefont {T.~E.}\ \bibnamefont {Northup}},\
  }\bibfield  {title} {\bibinfo {title} {{Entanglement} of {Trapped}-{Ion}
  {Qubits} {Separated} by 230 {Meters}},\ }\href
  {https://doi.org/10.1103/PhysRevLett.130.050803} {\bibfield  {journal}
  {\bibinfo  {journal} {Phys. Rev. Lett.}\ }\textbf {\bibinfo {volume} {130}},\
  \bibinfo {pages} {050803} (\bibinfo {year} {2023})}\BibitemShut {NoStop}%
\bibitem [{\citenamefont {Krutyanskiy}\ \emph {et~al.}(2022)\citenamefont
  {Krutyanskiy}, \citenamefont {Canteri}, \citenamefont {Meraner},
  \citenamefont {Bate}, \citenamefont {Krcmarsky}, \citenamefont {Schupp},
  \citenamefont {Sangouard},\ and\ \citenamefont
  {Lanyon}}]{krutyanskiy_telecom-wavelength_2022}%
  \BibitemOpen
  \bibfield  {author} {\bibinfo {author} {\bibfnamefont {V.}~\bibnamefont
  {Krutyanskiy}}, \bibinfo {author} {\bibfnamefont {M.}~\bibnamefont
  {Canteri}}, \bibinfo {author} {\bibfnamefont {M.}~\bibnamefont {Meraner}},
  \bibinfo {author} {\bibfnamefont {J.}~\bibnamefont {Bate}}, \bibinfo {author}
  {\bibfnamefont {V.}~\bibnamefont {Krcmarsky}}, \bibinfo {author}
  {\bibfnamefont {J.}~\bibnamefont {Schupp}}, \bibinfo {author} {\bibfnamefont
  {N.}~\bibnamefont {Sangouard}},\ and\ \bibinfo {author} {\bibfnamefont
  {B.~P.}\ \bibnamefont {Lanyon}},\ }\href {https://arxiv.org/abs/2210.05418}
  {\bibinfo {title} {A telecom-wavelength quantum repeater node based on a
  trapped-ion processor}} (\bibinfo {year} {2022}),\ \bibinfo {note}
  {arXiv:2210.05418}\BibitemShut {NoStop}%
\bibitem [{\citenamefont {Wang}\ \emph {et~al.}(2021)\citenamefont {Wang},
  \citenamefont {Luan}, \citenamefont {Qiao}, \citenamefont {Um}, \citenamefont
  {Zhang}, \citenamefont {Wang}, \citenamefont {Yuan}, \citenamefont {Gu},
  \citenamefont {Zhang},\ and\ \citenamefont {Kim}}]{wang_single_2021}%
  \BibitemOpen
  \bibfield  {author} {\bibinfo {author} {\bibfnamefont {P.}~\bibnamefont
  {Wang}}, \bibinfo {author} {\bibfnamefont {C.-Y.}\ \bibnamefont {Luan}},
  \bibinfo {author} {\bibfnamefont {M.}~\bibnamefont {Qiao}}, \bibinfo {author}
  {\bibfnamefont {M.}~\bibnamefont {Um}}, \bibinfo {author} {\bibfnamefont
  {J.}~\bibnamefont {Zhang}}, \bibinfo {author} {\bibfnamefont
  {Y.}~\bibnamefont {Wang}}, \bibinfo {author} {\bibfnamefont {X.}~\bibnamefont
  {Yuan}}, \bibinfo {author} {\bibfnamefont {M.}~\bibnamefont {Gu}}, \bibinfo
  {author} {\bibfnamefont {J.}~\bibnamefont {Zhang}},\ and\ \bibinfo {author}
  {\bibfnamefont {K.}~\bibnamefont {Kim}},\ }\bibfield  {title} {\bibinfo
  {title} {Single ion qubit with estimated coherence time exceeding one hour},\
  }\href {http://www.nature.com/articles/s41467-020-20330-w} {\bibfield
  {journal} {\bibinfo  {journal} {Nat. Commun.}\ }\textbf {\bibinfo {volume}
  {12}},\ \bibinfo {pages} {233} (\bibinfo {year} {2021})}\BibitemShut
  {NoStop}%
\bibitem [{\citenamefont {Christensen}\ \emph {et~al.}(2020)\citenamefont
  {Christensen}, \citenamefont {Hucul}, \citenamefont {Campbell},\ and\
  \citenamefont {Hudson}}]{christensen_high-fidelity_2020}%
  \BibitemOpen
  \bibfield  {author} {\bibinfo {author} {\bibfnamefont {J.~E.}\ \bibnamefont
  {Christensen}}, \bibinfo {author} {\bibfnamefont {D.}~\bibnamefont {Hucul}},
  \bibinfo {author} {\bibfnamefont {W.~C.}\ \bibnamefont {Campbell}},\ and\
  \bibinfo {author} {\bibfnamefont {E.~R.}\ \bibnamefont {Hudson}},\ }\bibfield
   {title} {\bibinfo {title} {High-fidelity manipulation of a qubit enabled by
  a manufactured nucleus},\ }\href
  {http://www.nature.com/articles/s41534-020-0265-5} {\bibfield  {journal}
  {\bibinfo  {journal} {npj Quantum Inf.}\ }\textbf {\bibinfo {volume} {6}},\
  \bibinfo {pages} {35} (\bibinfo {year} {2020})}\BibitemShut {NoStop}%
\bibitem [{\citenamefont {Harty}\ \emph {et~al.}(2014)\citenamefont {Harty},
  \citenamefont {Allcock}, \citenamefont {Ballance}, \citenamefont {Guidoni},
  \citenamefont {Janacek}, \citenamefont {Linke}, \citenamefont {Stacey},\ and\
  \citenamefont {Lucas}}]{harty_high-fidelity_2014}%
  \BibitemOpen
  \bibfield  {author} {\bibinfo {author} {\bibfnamefont {T.~P.}\ \bibnamefont
  {Harty}}, \bibinfo {author} {\bibfnamefont {D.~T.~C.}\ \bibnamefont
  {Allcock}}, \bibinfo {author} {\bibfnamefont {C.~J.}\ \bibnamefont
  {Ballance}}, \bibinfo {author} {\bibfnamefont {L.}~\bibnamefont {Guidoni}},
  \bibinfo {author} {\bibfnamefont {H.~A.}\ \bibnamefont {Janacek}}, \bibinfo
  {author} {\bibfnamefont {N.~M.}\ \bibnamefont {Linke}}, \bibinfo {author}
  {\bibfnamefont {D.~N.}\ \bibnamefont {Stacey}},\ and\ \bibinfo {author}
  {\bibfnamefont {D.~M.}\ \bibnamefont {Lucas}},\ }\bibfield  {title} {\bibinfo
  {title} {High-fidelity preparation, gates, memory, and readout of a
  trapped-ion quantum bit},\ }\href
  {https://doi.org/10.1103/PhysRevLett.113.220501} {\bibfield  {journal}
  {\bibinfo  {journal} {Phys. Rev. Lett.}\ }\textbf {\bibinfo {volume} {113}},\
  \bibinfo {pages} {220501} (\bibinfo {year} {2014})}\BibitemShut {NoStop}%
\bibitem [{\citenamefont {Srinivas}\ \emph {et~al.}(2021)\citenamefont
  {Srinivas}, \citenamefont {Burd}, \citenamefont {Knaack}, \citenamefont
  {Sutherland}, \citenamefont {Kwiatkowski}, \citenamefont {Glancy},
  \citenamefont {Knill}, \citenamefont {Wineland}, \citenamefont {Leibfried},
  \citenamefont {Wilson}, \citenamefont {Allcock},\ and\ \citenamefont
  {Slichter}}]{srinivas_high-fidelity_2021}%
  \BibitemOpen
  \bibfield  {author} {\bibinfo {author} {\bibfnamefont {R.}~\bibnamefont
  {Srinivas}}, \bibinfo {author} {\bibfnamefont {S.~C.}\ \bibnamefont {Burd}},
  \bibinfo {author} {\bibfnamefont {H.~M.}\ \bibnamefont {Knaack}}, \bibinfo
  {author} {\bibfnamefont {R.~T.}\ \bibnamefont {Sutherland}}, \bibinfo
  {author} {\bibfnamefont {A.}~\bibnamefont {Kwiatkowski}}, \bibinfo {author}
  {\bibfnamefont {S.}~\bibnamefont {Glancy}}, \bibinfo {author} {\bibfnamefont
  {E.}~\bibnamefont {Knill}}, \bibinfo {author} {\bibfnamefont {D.~J.}\
  \bibnamefont {Wineland}}, \bibinfo {author} {\bibfnamefont {D.}~\bibnamefont
  {Leibfried}}, \bibinfo {author} {\bibfnamefont {A.~C.}\ \bibnamefont
  {Wilson}}, \bibinfo {author} {\bibfnamefont {D.~T.~C.}\ \bibnamefont
  {Allcock}},\ and\ \bibinfo {author} {\bibfnamefont {D.~H.}\ \bibnamefont
  {Slichter}},\ }\bibfield  {title} {\bibinfo {title} {High-fidelity laser-free
  universal control of trapped ion qubits},\ }\href
  {https://www.nature.com/articles/s41586-021-03809-4} {\bibfield  {journal}
  {\bibinfo  {journal} {Nature}\ }\textbf {\bibinfo {volume} {597}},\ \bibinfo
  {pages} {209} (\bibinfo {year} {2021})}\BibitemShut {NoStop}%
\bibitem [{\citenamefont {Gaebler}\ \emph {et~al.}(2016)\citenamefont
  {Gaebler}, \citenamefont {Tan}, \citenamefont {Lin}, \citenamefont {Wan},
  \citenamefont {Bowler}, \citenamefont {Keith}, \citenamefont {Glancy},
  \citenamefont {Coakley}, \citenamefont {Knill}, \citenamefont {Leibfried},\
  and\ \citenamefont {Wineland}}]{gaebler_high-fidelity_2016}%
  \BibitemOpen
  \bibfield  {author} {\bibinfo {author} {\bibfnamefont {J.~P.}\ \bibnamefont
  {Gaebler}}, \bibinfo {author} {\bibfnamefont {T.~R.}\ \bibnamefont {Tan}},
  \bibinfo {author} {\bibfnamefont {Y.}~\bibnamefont {Lin}}, \bibinfo {author}
  {\bibfnamefont {Y.}~\bibnamefont {Wan}}, \bibinfo {author} {\bibfnamefont
  {R.}~\bibnamefont {Bowler}}, \bibinfo {author} {\bibfnamefont {A.~C.}\
  \bibnamefont {Keith}}, \bibinfo {author} {\bibfnamefont {S.}~\bibnamefont
  {Glancy}}, \bibinfo {author} {\bibfnamefont {K.}~\bibnamefont {Coakley}},
  \bibinfo {author} {\bibfnamefont {E.}~\bibnamefont {Knill}}, \bibinfo
  {author} {\bibfnamefont {D.}~\bibnamefont {Leibfried}},\ and\ \bibinfo
  {author} {\bibfnamefont {D.~J.}\ \bibnamefont {Wineland}},\ }\bibfield
  {title} {\bibinfo {title} {High-fidelity universal gate set for
  ${^{9}\mathrm{Be}}^{+}$ ion qubits},\ }\href
  {https://doi.org/10.1103/PhysRevLett.117.060505} {\bibfield  {journal}
  {\bibinfo  {journal} {Phys. Rev. Lett.}\ }\textbf {\bibinfo {volume} {117}},\
  \bibinfo {pages} {060505} (\bibinfo {year} {2016})}\BibitemShut {NoStop}%
\bibitem [{\citenamefont {Clark}\ \emph {et~al.}(2021)\citenamefont {Clark},
  \citenamefont {Tinkey}, \citenamefont {Sawyer}, \citenamefont {Meier},
  \citenamefont {Burkhardt}, \citenamefont {Seck}, \citenamefont {Shappert},
  \citenamefont {Guise}, \citenamefont {Volin}, \citenamefont {Fallek},
  \citenamefont {Hayden}, \citenamefont {Rellergert},\ and\ \citenamefont
  {Brown}}]{clark_high-fidelity_2021}%
  \BibitemOpen
  \bibfield  {author} {\bibinfo {author} {\bibfnamefont {C.~R.}\ \bibnamefont
  {Clark}}, \bibinfo {author} {\bibfnamefont {H.~N.}\ \bibnamefont {Tinkey}},
  \bibinfo {author} {\bibfnamefont {B.~C.}\ \bibnamefont {Sawyer}}, \bibinfo
  {author} {\bibfnamefont {A.~M.}\ \bibnamefont {Meier}}, \bibinfo {author}
  {\bibfnamefont {K.~A.}\ \bibnamefont {Burkhardt}}, \bibinfo {author}
  {\bibfnamefont {C.~M.}\ \bibnamefont {Seck}}, \bibinfo {author}
  {\bibfnamefont {C.~M.}\ \bibnamefont {Shappert}}, \bibinfo {author}
  {\bibfnamefont {N.~D.}\ \bibnamefont {Guise}}, \bibinfo {author}
  {\bibfnamefont {C.~E.}\ \bibnamefont {Volin}}, \bibinfo {author}
  {\bibfnamefont {S.~D.}\ \bibnamefont {Fallek}}, \bibinfo {author}
  {\bibfnamefont {H.~T.}\ \bibnamefont {Hayden}}, \bibinfo {author}
  {\bibfnamefont {W.~G.}\ \bibnamefont {Rellergert}},\ and\ \bibinfo {author}
  {\bibfnamefont {K.~R.}\ \bibnamefont {Brown}},\ }\bibfield  {title} {\bibinfo
  {title} {High-fidelity bell-state preparation with $^{40}\mathrm{Ca}^{+}$
  optical qubits},\ }\href {https://doi.org/10.1103/PhysRevLett.127.130505}
  {\bibfield  {journal} {\bibinfo  {journal} {Phys. Rev. Lett.}\ }\textbf
  {\bibinfo {volume} {127}},\ \bibinfo {pages} {130505} (\bibinfo {year}
  {2021})}\BibitemShut {NoStop}%
\bibitem [{\citenamefont {Blinov}\ \emph {et~al.}(2004)\citenamefont {Blinov},
  \citenamefont {Moehring}, \citenamefont {Duan},\ and\ \citenamefont
  {Monroe}}]{blinov_observation_2004}%
  \BibitemOpen
  \bibfield  {author} {\bibinfo {author} {\bibfnamefont {B.~B.}\ \bibnamefont
  {Blinov}}, \bibinfo {author} {\bibfnamefont {D.~L.}\ \bibnamefont
  {Moehring}}, \bibinfo {author} {\bibfnamefont {L.-M.}\ \bibnamefont {Duan}},\
  and\ \bibinfo {author} {\bibfnamefont {C.}~\bibnamefont {Monroe}},\
  }\bibfield  {title} {\bibinfo {title} {Observation of entanglement between a
  single trapped atom and a single photon},\ }\href
  {https://www.nature.com/articles/nature02377} {\bibfield  {journal} {\bibinfo
   {journal} {Nature}\ }\textbf {\bibinfo {volume} {428}},\ \bibinfo {pages}
  {153} (\bibinfo {year} {2004})}\BibitemShut {NoStop}%
\bibitem [{\citenamefont {Yang}\ \emph {et~al.}(2022)\citenamefont {Yang},
  \citenamefont {Ma}, \citenamefont {Wu}, \citenamefont {Wang}, \citenamefont
  {Cao}, \citenamefont {Guo}, \citenamefont {Huang}, \citenamefont {Feng},
  \citenamefont {Zhou},\ and\ \citenamefont {Duan}}]{yang_realizing_2022}%
  \BibitemOpen
  \bibfield  {author} {\bibinfo {author} {\bibfnamefont {H.-X.}\ \bibnamefont
  {Yang}}, \bibinfo {author} {\bibfnamefont {J.-Y.}\ \bibnamefont {Ma}},
  \bibinfo {author} {\bibfnamefont {Y.-K.}\ \bibnamefont {Wu}}, \bibinfo
  {author} {\bibfnamefont {Y.}~\bibnamefont {Wang}}, \bibinfo {author}
  {\bibfnamefont {M.-M.}\ \bibnamefont {Cao}}, \bibinfo {author} {\bibfnamefont
  {W.-X.}\ \bibnamefont {Guo}}, \bibinfo {author} {\bibfnamefont {Y.-Y.}\
  \bibnamefont {Huang}}, \bibinfo {author} {\bibfnamefont {L.}~\bibnamefont
  {Feng}}, \bibinfo {author} {\bibfnamefont {Z.-C.}\ \bibnamefont {Zhou}},\
  and\ \bibinfo {author} {\bibfnamefont {L.-M.}\ \bibnamefont {Duan}},\
  }\bibfield  {title} {\bibinfo {title} {Realizing coherently convertible
  dual-type qubits with the same ion species},\ }\href
  {https://doi.org/10.1038/s41567-022-01661-5} {\bibfield  {journal} {\bibinfo
  {journal} {Nat. Phys.}\ }\textbf {\bibinfo {volume} {18}},\ \bibinfo {pages}
  {1058} (\bibinfo {year} {2022})}\BibitemShut {NoStop}%
\bibitem [{\citenamefont {Inlek}\ \emph {et~al.}(2017)\citenamefont {Inlek},
  \citenamefont {Crocker}, \citenamefont {Lichtman}, \citenamefont {Sosnova},\
  and\ \citenamefont {Monroe}}]{inlek_multispecies_2017}%
  \BibitemOpen
  \bibfield  {author} {\bibinfo {author} {\bibfnamefont {I.~V.}\ \bibnamefont
  {Inlek}}, \bibinfo {author} {\bibfnamefont {C.}~\bibnamefont {Crocker}},
  \bibinfo {author} {\bibfnamefont {M.}~\bibnamefont {Lichtman}}, \bibinfo
  {author} {\bibfnamefont {K.}~\bibnamefont {Sosnova}},\ and\ \bibinfo {author}
  {\bibfnamefont {C.}~\bibnamefont {Monroe}},\ }\bibfield  {title} {\bibinfo
  {title} {Multispecies trapped-ion node for quantum networking},\ }\href
  {https://doi.org/10.1103/PhysRevLett.118.250502} {\bibfield  {journal}
  {\bibinfo  {journal} {Phys. Rev. Lett.}\ }\textbf {\bibinfo {volume} {118}},\
  \bibinfo {pages} {250502} (\bibinfo {year} {2017})}\BibitemShut {NoStop}%
\bibitem [{\citenamefont {Negnevitsky}\ \emph {et~al.}(2018)\citenamefont
  {Negnevitsky}, \citenamefont {Marinelli}, \citenamefont {Mehta},
  \citenamefont {Lo}, \citenamefont {Fl{\"{u}}hmann},\ and\ \citenamefont
  {Home}}]{negnevitsky_repeated_2018}%
  \BibitemOpen
  \bibfield  {author} {\bibinfo {author} {\bibfnamefont {V.}~\bibnamefont
  {Negnevitsky}}, \bibinfo {author} {\bibfnamefont {M.}~\bibnamefont
  {Marinelli}}, \bibinfo {author} {\bibfnamefont {K.~K.}\ \bibnamefont
  {Mehta}}, \bibinfo {author} {\bibfnamefont {H.-Y.}\ \bibnamefont {Lo}},
  \bibinfo {author} {\bibfnamefont {C.}~\bibnamefont {Fl{\"{u}}hmann}},\ and\
  \bibinfo {author} {\bibfnamefont {J.~P.}\ \bibnamefont {Home}},\ }\bibfield
  {title} {\bibinfo {title} {Repeated multi-qubit readout and feedback with a
  mixed-species trapped-ion register},\ }\href
  {http://www.nature.com/articles/s41586-018-0668-z} {\bibfield  {journal}
  {\bibinfo  {journal} {Nature}\ }\textbf {\bibinfo {volume} {563}},\ \bibinfo
  {pages} {527} (\bibinfo {year} {2018})}\BibitemShut {NoStop}%
\bibitem [{\citenamefont {Lucas}\ \emph {et~al.}(2007)\citenamefont {Lucas},
  \citenamefont {Keitch}, \citenamefont {Home}, \citenamefont {Imreh},
  \citenamefont {McDonnell}, \citenamefont {Stacey}, \citenamefont {Szwer},\
  and\ \citenamefont {Steane}}]{lucas_long-lived_2007}%
  \BibitemOpen
  \bibfield  {author} {\bibinfo {author} {\bibfnamefont {D.~M.}\ \bibnamefont
  {Lucas}}, \bibinfo {author} {\bibfnamefont {B.~C.}\ \bibnamefont {Keitch}},
  \bibinfo {author} {\bibfnamefont {J.~P.}\ \bibnamefont {Home}}, \bibinfo
  {author} {\bibfnamefont {G.}~\bibnamefont {Imreh}}, \bibinfo {author}
  {\bibfnamefont {M.~J.}\ \bibnamefont {McDonnell}}, \bibinfo {author}
  {\bibfnamefont {D.~N.}\ \bibnamefont {Stacey}}, \bibinfo {author}
  {\bibfnamefont {D.~J.}\ \bibnamefont {Szwer}},\ and\ \bibinfo {author}
  {\bibfnamefont {A.~M.}\ \bibnamefont {Steane}},\ }\href
  {http://arxiv.org/abs/0710.4421} {\bibinfo {title} {A long-lived memory qubit
  on a low-decoherence quantum bus}} (\bibinfo {year} {2007}),\ \bibinfo {note}
  {arXiv:0710.4421}\BibitemShut {NoStop}%
\bibitem [{Note1()}]{Note1}%
  \BibitemOpen
  \bibinfo {note} {Sandia National Laboratories HOA2.}\BibitemShut {Stop}%
\bibitem [{sup()}]{supplement}%
  \BibitemOpen
  \href@noop {} {\bibinfo {title} {{See Supplemental Material}}}\BibitemShut
  {NoStop}%
\bibitem [{\citenamefont {Home}\ \emph {et~al.}(2009)\citenamefont {Home},
  \citenamefont {McDonnell}, \citenamefont {Szwer}, \citenamefont {Keitch},
  \citenamefont {Lucas}, \citenamefont {Stacey},\ and\ \citenamefont
  {Steane}}]{home_memory_2009}%
  \BibitemOpen
  \bibfield  {author} {\bibinfo {author} {\bibfnamefont {J.~P.}\ \bibnamefont
  {Home}}, \bibinfo {author} {\bibfnamefont {M.~J.}\ \bibnamefont {McDonnell}},
  \bibinfo {author} {\bibfnamefont {D.~J.}\ \bibnamefont {Szwer}}, \bibinfo
  {author} {\bibfnamefont {B.~C.}\ \bibnamefont {Keitch}}, \bibinfo {author}
  {\bibfnamefont {D.~M.}\ \bibnamefont {Lucas}}, \bibinfo {author}
  {\bibfnamefont {D.~N.}\ \bibnamefont {Stacey}},\ and\ \bibinfo {author}
  {\bibfnamefont {A.~M.}\ \bibnamefont {Steane}},\ }\bibfield  {title}
  {\bibinfo {title} {Memory coherence of a sympathetically cooled trapped-ion
  qubit},\ }\href {https://doi.org/10.1103/PhysRevA.79.050305} {\bibfield
  {journal} {\bibinfo  {journal} {Phys. Rev. A}\ }\textbf {\bibinfo {volume}
  {79}},\ \bibinfo {pages} {050305(R)} (\bibinfo {year} {2009})}\BibitemShut
  {NoStop}%
\bibitem [{\citenamefont {Ballance}\ \emph {et~al.}(2016)\citenamefont
  {Ballance}, \citenamefont {Harty}, \citenamefont {Linke}, \citenamefont
  {Sepiol},\ and\ \citenamefont {Lucas}}]{ballance_high-fidelity_2016}%
  \BibitemOpen
  \bibfield  {author} {\bibinfo {author} {\bibfnamefont {C.~J.}\ \bibnamefont
  {Ballance}}, \bibinfo {author} {\bibfnamefont {T.~P.}\ \bibnamefont {Harty}},
  \bibinfo {author} {\bibfnamefont {N.~M.}\ \bibnamefont {Linke}}, \bibinfo
  {author} {\bibfnamefont {M.~A.}\ \bibnamefont {Sepiol}},\ and\ \bibinfo
  {author} {\bibfnamefont {D.~M.}\ \bibnamefont {Lucas}},\ }\bibfield  {title}
  {\bibinfo {title} {High-fidelity quantum logic gates using trapped-ion
  hyperfine qubits},\ }\href {https://doi.org/10.1103/PhysRevLett.117.060504}
  {\bibfield  {journal} {\bibinfo  {journal} {Phys. Rev. Lett.}\ }\textbf
  {\bibinfo {volume} {117}},\ \bibinfo {pages} {060504} (\bibinfo {year}
  {2016})}\BibitemShut {NoStop}%
\bibitem [{\citenamefont {Sepiol}\ \emph {et~al.}(2019)\citenamefont {Sepiol},
  \citenamefont {Hughes}, \citenamefont {Tarlton}, \citenamefont {Nadlinger},
  \citenamefont {Ballance}, \citenamefont {Ballance}, \citenamefont {Harty},
  \citenamefont {Steane}, \citenamefont {Goodwin},\ and\ \citenamefont
  {Lucas}}]{sepiol_probing_2019}%
  \BibitemOpen
  \bibfield  {author} {\bibinfo {author} {\bibfnamefont {M.~A.}\ \bibnamefont
  {Sepiol}}, \bibinfo {author} {\bibfnamefont {A.~C.}\ \bibnamefont {Hughes}},
  \bibinfo {author} {\bibfnamefont {J.~E.}\ \bibnamefont {Tarlton}}, \bibinfo
  {author} {\bibfnamefont {D.~P.}\ \bibnamefont {Nadlinger}}, \bibinfo {author}
  {\bibfnamefont {T.~G.}\ \bibnamefont {Ballance}}, \bibinfo {author}
  {\bibfnamefont {C.~J.}\ \bibnamefont {Ballance}}, \bibinfo {author}
  {\bibfnamefont {T.~P.}\ \bibnamefont {Harty}}, \bibinfo {author}
  {\bibfnamefont {A.~M.}\ \bibnamefont {Steane}}, \bibinfo {author}
  {\bibfnamefont {J.~F.}\ \bibnamefont {Goodwin}},\ and\ \bibinfo {author}
  {\bibfnamefont {D.~M.}\ \bibnamefont {Lucas}},\ }\bibfield  {title} {\bibinfo
  {title} {Probing qubit memory errors at the part-per-million level},\ }\href
  {https://doi.org/10.1103/PhysRevLett.123.110503} {\bibfield  {journal}
  {\bibinfo  {journal} {Phys. Rev. Lett.}\ }\textbf {\bibinfo {volume} {123}},\
  \bibinfo {pages} {110503} (\bibinfo {year} {2019})}\BibitemShut {NoStop}%
\bibitem [{\citenamefont {Kielpinski}\ \emph {et~al.}(2000)\citenamefont
  {Kielpinski}, \citenamefont {King}, \citenamefont {Myatt}, \citenamefont
  {Sackett}, \citenamefont {Turchette}, \citenamefont {Itano}, \citenamefont
  {Monroe}, \citenamefont {Wineland},\ and\ \citenamefont
  {Zurek}}]{kielpinski_sympathetic_2000}%
  \BibitemOpen
  \bibfield  {author} {\bibinfo {author} {\bibfnamefont {D.}~\bibnamefont
  {Kielpinski}}, \bibinfo {author} {\bibfnamefont {B.~E.}\ \bibnamefont
  {King}}, \bibinfo {author} {\bibfnamefont {C.~J.}\ \bibnamefont {Myatt}},
  \bibinfo {author} {\bibfnamefont {C.~A.}\ \bibnamefont {Sackett}}, \bibinfo
  {author} {\bibfnamefont {Q.~A.}\ \bibnamefont {Turchette}}, \bibinfo {author}
  {\bibfnamefont {W.~M.}\ \bibnamefont {Itano}}, \bibinfo {author}
  {\bibfnamefont {C.}~\bibnamefont {Monroe}}, \bibinfo {author} {\bibfnamefont
  {D.~J.}\ \bibnamefont {Wineland}},\ and\ \bibinfo {author} {\bibfnamefont
  {W.~H.}\ \bibnamefont {Zurek}},\ }\bibfield  {title} {\bibinfo {title}
  {Sympathetic cooling of trapped ions for quantum logic},\ }\href
  {https://doi.org/10.1103/PhysRevA.61.032310} {\bibfield  {journal} {\bibinfo
  {journal} {Phys. Rev. A}\ }\textbf {\bibinfo {volume} {61}},\ \bibinfo
  {pages} {032310} (\bibinfo {year} {2000})}\BibitemShut {NoStop}%
\bibitem [{\citenamefont {Hughes}\ \emph {et~al.}(2020)\citenamefont {Hughes},
  \citenamefont {Sch\"afer}, \citenamefont {Thirumalai}, \citenamefont
  {Nadlinger}, \citenamefont {Woodrow}, \citenamefont {Lucas},\ and\
  \citenamefont {Ballance}}]{hughes_benchmarking_2020}%
  \BibitemOpen
  \bibfield  {author} {\bibinfo {author} {\bibfnamefont {A.~C.}\ \bibnamefont
  {Hughes}}, \bibinfo {author} {\bibfnamefont {V.~M.}\ \bibnamefont
  {Sch\"afer}}, \bibinfo {author} {\bibfnamefont {K.}~\bibnamefont
  {Thirumalai}}, \bibinfo {author} {\bibfnamefont {D.~P.}\ \bibnamefont
  {Nadlinger}}, \bibinfo {author} {\bibfnamefont {S.~R.}\ \bibnamefont
  {Woodrow}}, \bibinfo {author} {\bibfnamefont {D.~M.}\ \bibnamefont {Lucas}},\
  and\ \bibinfo {author} {\bibfnamefont {C.~J.}\ \bibnamefont {Ballance}},\
  }\bibfield  {title} {\bibinfo {title} {Benchmarking a high-fidelity
  mixed-species entangling gate},\ }\href
  {https://doi.org/10.1103/PhysRevLett.125.080504} {\bibfield  {journal}
  {\bibinfo  {journal} {Phys. Rev. Lett.}\ }\textbf {\bibinfo {volume} {125}},\
  \bibinfo {pages} {080504} (\bibinfo {year} {2020})}\BibitemShut {NoStop}%
\bibitem [{\citenamefont {Myerson}\ \emph {et~al.}(2008)\citenamefont
  {Myerson}, \citenamefont {Szwer}, \citenamefont {Webster}, \citenamefont
  {Allcock}, \citenamefont {Curtis}, \citenamefont {Imreh}, \citenamefont
  {Sherman}, \citenamefont {Stacey}, \citenamefont {Steane},\ and\
  \citenamefont {Lucas}}]{myerson_high-fidelity_2008}%
  \BibitemOpen
  \bibfield  {author} {\bibinfo {author} {\bibfnamefont {A.~H.}\ \bibnamefont
  {Myerson}}, \bibinfo {author} {\bibfnamefont {D.~J.}\ \bibnamefont {Szwer}},
  \bibinfo {author} {\bibfnamefont {S.~C.}\ \bibnamefont {Webster}}, \bibinfo
  {author} {\bibfnamefont {D.~T.~C.}\ \bibnamefont {Allcock}}, \bibinfo
  {author} {\bibfnamefont {M.~J.}\ \bibnamefont {Curtis}}, \bibinfo {author}
  {\bibfnamefont {G.}~\bibnamefont {Imreh}}, \bibinfo {author} {\bibfnamefont
  {J.~A.}\ \bibnamefont {Sherman}}, \bibinfo {author} {\bibfnamefont {D.~N.}\
  \bibnamefont {Stacey}}, \bibinfo {author} {\bibfnamefont {A.~M.}\
  \bibnamefont {Steane}},\ and\ \bibinfo {author} {\bibfnamefont {D.~M.}\
  \bibnamefont {Lucas}},\ }\bibfield  {title} {\bibinfo {title} {High-fidelity
  readout of trapped-ion qubits},\ }\href
  {https://doi.org/10.1103/PhysRevLett.100.200502} {\bibfield  {journal}
  {\bibinfo  {journal} {Phys. Rev. Lett.}\ }\textbf {\bibinfo {volume} {100}},\
  \bibinfo {pages} {200502} (\bibinfo {year} {2008})}\BibitemShut {NoStop}%
\bibitem [{\citenamefont {Main}(2020)}]{main_magnetic_2020}%
  \BibitemOpen
  \bibfield  {author} {\bibinfo {author} {\bibfnamefont {D.}~\bibnamefont
  {Main}},\ }\emph {\bibinfo {title} {Magnetic {Field} {Stabilisation} in {Ion}
  {Traps}}},\ \href@noop {} {Master's thesis},\ \bibinfo  {school} {University
  of Oxford} (\bibinfo {year} {2020})\BibitemShut {NoStop}%
\bibitem [{\citenamefont {Bruzewicz}\ \emph {et~al.}(2019)\citenamefont
  {Bruzewicz}, \citenamefont {McConnell}, \citenamefont {Stuart}, \citenamefont
  {Sage},\ and\ \citenamefont {Chiaverini}}]{bruzewicz_dual-species_2019}%
  \BibitemOpen
  \bibfield  {author} {\bibinfo {author} {\bibfnamefont {C.~D.}\ \bibnamefont
  {Bruzewicz}}, \bibinfo {author} {\bibfnamefont {R.}~\bibnamefont
  {McConnell}}, \bibinfo {author} {\bibfnamefont {J.}~\bibnamefont {Stuart}},
  \bibinfo {author} {\bibfnamefont {J.~M.}\ \bibnamefont {Sage}},\ and\
  \bibinfo {author} {\bibfnamefont {J.}~\bibnamefont {Chiaverini}},\ }\bibfield
   {title} {\bibinfo {title} {Dual-species, multi-qubit logic primitives for
  $\mathrm{Ca}^{+}$/$\mathrm{Sr}^{+}$ trapped-ion crystals},\ }\href
  {http://www.nature.com/articles/s41534-019-0218-z} {\bibfield  {journal}
  {\bibinfo  {journal} {npj Quantum Inf.}\ }\textbf {\bibinfo {volume} {5}},\
  \bibinfo {pages} {102} (\bibinfo {year} {2019})}\BibitemShut {NoStop}%
\bibitem [{\citenamefont {Nadlinger}\ \emph {et~al.}(2021)\citenamefont
  {Nadlinger}, \citenamefont {Drmota}, \citenamefont {Main}, \citenamefont
  {Nichol}, \citenamefont {Araneda}, \citenamefont {Srinivas}, \citenamefont
  {Stephenson}, \citenamefont {Ballance},\ and\ \citenamefont
  {Lucas}}]{nadlinger_micromotion_2021}%
  \BibitemOpen
  \bibfield  {author} {\bibinfo {author} {\bibfnamefont {D.~P.}\ \bibnamefont
  {Nadlinger}}, \bibinfo {author} {\bibfnamefont {P.}~\bibnamefont {Drmota}},
  \bibinfo {author} {\bibfnamefont {D.}~\bibnamefont {Main}}, \bibinfo {author}
  {\bibfnamefont {B.~C.}\ \bibnamefont {Nichol}}, \bibinfo {author}
  {\bibfnamefont {G.}~\bibnamefont {Araneda}}, \bibinfo {author} {\bibfnamefont
  {R.}~\bibnamefont {Srinivas}}, \bibinfo {author} {\bibfnamefont {L.~J.}\
  \bibnamefont {Stephenson}}, \bibinfo {author} {\bibfnamefont {C.~J.}\
  \bibnamefont {Ballance}},\ and\ \bibinfo {author} {\bibfnamefont {D.~M.}\
  \bibnamefont {Lucas}},\ }\href {https://arxiv.org/abs/2107.00056} {\bibinfo
  {title} {Micromotion minimisation by synchronous detection of parametrically
  excited motion}} (\bibinfo {year} {2021}),\ \bibinfo {note}
  {arXiv:2107.00056}\BibitemShut {NoStop}%
\bibitem [{\citenamefont {Stas}\ \emph {et~al.}(2022)\citenamefont {Stas},
  \citenamefont {Huan}, \citenamefont {Machielse}, \citenamefont {Knall},
  \citenamefont {Suleymanzade}, \citenamefont {Pingault}, \citenamefont
  {Sutula}, \citenamefont {Ding}, \citenamefont {Knaut}, \citenamefont
  {Assumpcao}, \citenamefont {Wei}, \citenamefont {Bhaskar}, \citenamefont
  {Riedinger}, \citenamefont {Sukachev}, \citenamefont {Park}, \citenamefont
  {Lon{\v{c}}ar}, \citenamefont {Levonian},\ and\ \citenamefont
  {Lukin}}]{stas_robust_2022}%
  \BibitemOpen
  \bibfield  {author} {\bibinfo {author} {\bibfnamefont {P.-J.}\ \bibnamefont
  {Stas}}, \bibinfo {author} {\bibfnamefont {Y.~Q.}\ \bibnamefont {Huan}},
  \bibinfo {author} {\bibfnamefont {B.}~\bibnamefont {Machielse}}, \bibinfo
  {author} {\bibfnamefont {E.~N.}\ \bibnamefont {Knall}}, \bibinfo {author}
  {\bibfnamefont {A.}~\bibnamefont {Suleymanzade}}, \bibinfo {author}
  {\bibfnamefont {B.}~\bibnamefont {Pingault}}, \bibinfo {author}
  {\bibfnamefont {M.}~\bibnamefont {Sutula}}, \bibinfo {author} {\bibfnamefont
  {S.~W.}\ \bibnamefont {Ding}}, \bibinfo {author} {\bibfnamefont {C.~M.}\
  \bibnamefont {Knaut}}, \bibinfo {author} {\bibfnamefont {D.~R.}\ \bibnamefont
  {Assumpcao}}, \bibinfo {author} {\bibfnamefont {Y.-C.}\ \bibnamefont {Wei}},
  \bibinfo {author} {\bibfnamefont {M.~K.}\ \bibnamefont {Bhaskar}}, \bibinfo
  {author} {\bibfnamefont {R.}~\bibnamefont {Riedinger}}, \bibinfo {author}
  {\bibfnamefont {D.~D.}\ \bibnamefont {Sukachev}}, \bibinfo {author}
  {\bibfnamefont {H.}~\bibnamefont {Park}}, \bibinfo {author} {\bibfnamefont
  {M.}~\bibnamefont {Lon{\v{c}}ar}}, \bibinfo {author} {\bibfnamefont {D.~S.}\
  \bibnamefont {Levonian}},\ and\ \bibinfo {author} {\bibfnamefont {M.~D.}\
  \bibnamefont {Lukin}},\ }\bibfield  {title} {\bibinfo {title} {Robust
  multi-qubit quantum network node with integrated error detection},\ }\href
  {https://www.science.org/doi/abs/10.1126/science.add9771} {\bibfield
  {journal} {\bibinfo  {journal} {Science}\ }\textbf {\bibinfo {volume}
  {378}},\ \bibinfo {pages} {557} (\bibinfo {year} {2022})}\BibitemShut
  {NoStop}%
\bibitem [{\citenamefont {\ifmmode \check{R}\else
  \v{R}\fi{}eh\'a\ifmmode~\check{c}\else \v{c}\fi{}ek}\ \emph
  {et~al.}(2007)\citenamefont {\ifmmode \check{R}\else
  \v{R}\fi{}eh\'a\ifmmode~\check{c}\else \v{c}\fi{}ek}, \citenamefont {Hradil},
  \citenamefont {Knill},\ and\ \citenamefont {Lvovsky}}]{rehacek_diluted_2007}%
  \BibitemOpen
  \bibfield  {author} {\bibinfo {author} {\bibfnamefont {J.}~\bibnamefont
  {\ifmmode \check{R}\else \v{R}\fi{}eh\'a\ifmmode~\check{c}\else
  \v{c}\fi{}ek}}, \bibinfo {author} {\bibfnamefont {Z.~c.~v.}\ \bibnamefont
  {Hradil}}, \bibinfo {author} {\bibfnamefont {E.}~\bibnamefont {Knill}},\ and\
  \bibinfo {author} {\bibfnamefont {A.~I.}\ \bibnamefont {Lvovsky}},\
  }\bibfield  {title} {\bibinfo {title} {Diluted maximum-likelihood algorithm
  for quantum tomography},\ }\href {https://doi.org/10.1103/PhysRevA.75.042108}
  {\bibfield  {journal} {\bibinfo  {journal} {Phys. Rev. A}\ }\textbf {\bibinfo
  {volume} {75}},\ \bibinfo {pages} {042108} (\bibinfo {year}
  {2007})}\BibitemShut {NoStop}%
\bibitem [{\citenamefont {Anis}\ and\ \citenamefont
  {Lvovsky}(2012)}]{anis_maximum-likelihood_2012}%
  \BibitemOpen
  \bibfield  {author} {\bibinfo {author} {\bibfnamefont {A.}~\bibnamefont
  {Anis}}\ and\ \bibinfo {author} {\bibfnamefont {A.~I.}\ \bibnamefont
  {Lvovsky}},\ }\bibfield  {title} {\bibinfo {title} {Maximum-likelihood
  coherent-state quantum process tomography},\ }\href
  {https://iopscience.iop.org/article/10.1088/1367-2630/14/10/105021}
  {\bibfield  {journal} {\bibinfo  {journal} {New J. Phys.}\ }\textbf {\bibinfo
  {volume} {14}},\ \bibinfo {pages} {105021} (\bibinfo {year}
  {2012})}\BibitemShut {NoStop}%
\bibitem [{\citenamefont {Badzi{\c{a}}g}\ \emph {et~al.}(2000)\citenamefont
  {Badzi{\c{a}}g}, \citenamefont {Horodecki}, \citenamefont {Horodecki},\ and\
  \citenamefont {Horodecki}}]{badziag_local_2000}%
  \BibitemOpen
  \bibfield  {author} {\bibinfo {author} {\bibfnamefont {P.}~\bibnamefont
  {Badzi{\c{a}}g}}, \bibinfo {author} {\bibfnamefont {M.}~\bibnamefont
  {Horodecki}}, \bibinfo {author} {\bibfnamefont {P.}~\bibnamefont
  {Horodecki}},\ and\ \bibinfo {author} {\bibfnamefont {R.}~\bibnamefont
  {Horodecki}},\ }\bibfield  {title} {\bibinfo {title} {Local environment can
  enhance fidelity of quantum teleportation},\ }\href
  {https://doi.org/10.1103/PhysRevA.62.012311} {\bibfield  {journal} {\bibinfo
  {journal} {Phys. Rev. A}\ }\textbf {\bibinfo {volume} {62}},\ \bibinfo
  {pages} {012311} (\bibinfo {year} {2000})}\BibitemShut {NoStop}%
\bibitem [{\citenamefont {Souza}\ \emph {et~al.}(2011)\citenamefont {Souza},
  \citenamefont {\'Alvarez},\ and\ \citenamefont {Suter}}]{souza_robust_2011}%
  \BibitemOpen
  \bibfield  {author} {\bibinfo {author} {\bibfnamefont {A.~M.}\ \bibnamefont
  {Souza}}, \bibinfo {author} {\bibfnamefont {G.~A.}\ \bibnamefont
  {\'Alvarez}},\ and\ \bibinfo {author} {\bibfnamefont {D.}~\bibnamefont
  {Suter}},\ }\bibfield  {title} {\bibinfo {title} {Robust dynamical decoupling
  for quantum computing and quantum memory},\ }\href
  {https://doi.org/10.1103/PhysRevLett.106.240501} {\bibfield  {journal}
  {\bibinfo  {journal} {Phys. Rev. Lett.}\ }\textbf {\bibinfo {volume} {106}},\
  \bibinfo {pages} {240501} (\bibinfo {year} {2011})}\BibitemShut {NoStop}%
\bibitem [{\citenamefont {Monroe}\ \emph {et~al.}(2014)\citenamefont {Monroe},
  \citenamefont {Raussendorf}, \citenamefont {Ruthven}, \citenamefont {Brown},
  \citenamefont {Maunz}, \citenamefont {Duan},\ and\ \citenamefont
  {Kim}}]{monroe_large_2014}%
  \BibitemOpen
  \bibfield  {author} {\bibinfo {author} {\bibfnamefont {C.}~\bibnamefont
  {Monroe}}, \bibinfo {author} {\bibfnamefont {R.}~\bibnamefont {Raussendorf}},
  \bibinfo {author} {\bibfnamefont {A.}~\bibnamefont {Ruthven}}, \bibinfo
  {author} {\bibfnamefont {K.~R.}\ \bibnamefont {Brown}}, \bibinfo {author}
  {\bibfnamefont {P.}~\bibnamefont {Maunz}}, \bibinfo {author} {\bibfnamefont
  {L.-M.}\ \bibnamefont {Duan}},\ and\ \bibinfo {author} {\bibfnamefont
  {J.}~\bibnamefont {Kim}},\ }\bibfield  {title} {\bibinfo {title} {Large-scale
  modular quantum-computer architecture with atomic memory and photonic
  interconnects},\ }\href {https://doi.org/10.1103/PhysRevA.89.022317}
  {\bibfield  {journal} {\bibinfo  {journal} {Phys. Rev. A}\ }\textbf {\bibinfo
  {volume} {89}},\ \bibinfo {pages} {022317} (\bibinfo {year}
  {2014})}\BibitemShut {NoStop}%
\bibitem [{\citenamefont {Megidish}\ \emph {et~al.}(2013)\citenamefont
  {Megidish}, \citenamefont {Halevy}, \citenamefont {Shacham}, \citenamefont
  {Dvir}, \citenamefont {Dovrat},\ and\ \citenamefont
  {Eisenberg}}]{megidish_entanglement_2013}%
  \BibitemOpen
  \bibfield  {author} {\bibinfo {author} {\bibfnamefont {E.}~\bibnamefont
  {Megidish}}, \bibinfo {author} {\bibfnamefont {A.}~\bibnamefont {Halevy}},
  \bibinfo {author} {\bibfnamefont {T.}~\bibnamefont {Shacham}}, \bibinfo
  {author} {\bibfnamefont {T.}~\bibnamefont {Dvir}}, \bibinfo {author}
  {\bibfnamefont {L.}~\bibnamefont {Dovrat}},\ and\ \bibinfo {author}
  {\bibfnamefont {H.~S.}\ \bibnamefont {Eisenberg}},\ }\bibfield  {title}
  {\bibinfo {title} {Entanglement swapping between photons that have never
  coexisted},\ }\href {https://doi.org/10.1103/PhysRevLett.110.210403}
  {\bibfield  {journal} {\bibinfo  {journal} {Phys. Rev. Lett.}\ }\textbf
  {\bibinfo {volume} {110}},\ \bibinfo {pages} {210403} (\bibinfo {year}
  {2013})}\BibitemShut {NoStop}%
\bibitem [{\citenamefont {Bennett}\ \emph {et~al.}(1996)\citenamefont
  {Bennett}, \citenamefont {Brassard}, \citenamefont {Popescu}, \citenamefont
  {Schumacher}, \citenamefont {Smolin},\ and\ \citenamefont
  {Wootters}}]{bennett_purification_1996}%
  \BibitemOpen
  \bibfield  {author} {\bibinfo {author} {\bibfnamefont {C.~H.}\ \bibnamefont
  {Bennett}}, \bibinfo {author} {\bibfnamefont {G.}~\bibnamefont {Brassard}},
  \bibinfo {author} {\bibfnamefont {S.}~\bibnamefont {Popescu}}, \bibinfo
  {author} {\bibfnamefont {B.}~\bibnamefont {Schumacher}}, \bibinfo {author}
  {\bibfnamefont {J.~A.}\ \bibnamefont {Smolin}},\ and\ \bibinfo {author}
  {\bibfnamefont {W.~K.}\ \bibnamefont {Wootters}},\ }\bibfield  {title}
  {\bibinfo {title} {Purification of noisy entanglement and faithful
  teleportation via noisy channels},\ }\href
  {https://doi.org/10.1103/PhysRevLett.76.722} {\bibfield  {journal} {\bibinfo
  {journal} {Phys. Rev. Lett.}\ }\textbf {\bibinfo {volume} {76}},\ \bibinfo
  {pages} {722} (\bibinfo {year} {1996})}\BibitemShut {NoStop}%
\bibitem [{\citenamefont {Monroe}\ and\ \citenamefont
  {Kim}(2013)}]{monroe_scaling_2013}%
  \BibitemOpen
  \bibfield  {author} {\bibinfo {author} {\bibfnamefont {C.}~\bibnamefont
  {Monroe}}\ and\ \bibinfo {author} {\bibfnamefont {J.}~\bibnamefont {Kim}},\
  }\bibfield  {title} {\bibinfo {title} {Scaling the ion trap quantum
  processor},\ }\href {https://www.science.org/doi/abs/10.1126/science.1231298}
  {\bibfield  {journal} {\bibinfo  {journal} {Science}\ }\textbf {\bibinfo
  {volume} {339}},\ \bibinfo {pages} {1164} (\bibinfo {year}
  {2013})}\BibitemShut {NoStop}%
\bibitem [{\citenamefont {Nigmatullin}\ \emph {et~al.}(2016)\citenamefont
  {Nigmatullin}, \citenamefont {Ballance}, \citenamefont {Beaudrap},\ and\
  \citenamefont {Benjamin}}]{nigmatullin_minimally_2016}%
  \BibitemOpen
  \bibfield  {author} {\bibinfo {author} {\bibfnamefont {R.}~\bibnamefont
  {Nigmatullin}}, \bibinfo {author} {\bibfnamefont {C.~J.}\ \bibnamefont
  {Ballance}}, \bibinfo {author} {\bibfnamefont {N.~d.}\ \bibnamefont
  {Beaudrap}},\ and\ \bibinfo {author} {\bibfnamefont {S.~C.}\ \bibnamefont
  {Benjamin}},\ }\bibfield  {title} {\bibinfo {title} {Minimally complex ion
  traps as modules for quantum communication and computing},\ }\href
  {https://doi.org/10.1088%2F1367-2630%2F18%2F10%2F103028} {\bibfield
  {journal} {\bibinfo  {journal} {New J. Phys.}\ }\textbf {\bibinfo {volume}
  {18}},\ \bibinfo {pages} {103028} (\bibinfo {year} {2016})}\BibitemShut
  {NoStop}%
\bibitem [{\citenamefont {Broadbent}\ \emph {et~al.}(2009)\citenamefont
  {Broadbent}, \citenamefont {Fitzsimons},\ and\ \citenamefont
  {Kashefi}}]{broadbent_universal_2009}%
  \BibitemOpen
  \bibfield  {author} {\bibinfo {author} {\bibfnamefont {A.}~\bibnamefont
  {Broadbent}}, \bibinfo {author} {\bibfnamefont {J.}~\bibnamefont
  {Fitzsimons}},\ and\ \bibinfo {author} {\bibfnamefont {E.}~\bibnamefont
  {Kashefi}},\ }\bibfield  {title} {\bibinfo {title} {Universal blind quantum
  computation},\ }in\ \href {https://doi.org/10.1109/FOCS.2009.36} {\emph
  {\bibinfo {booktitle} {2009 50th Annual IEEE Symposium on Foundations of
  Computer Science}}}\ (\bibinfo {year} {2009})\ pp.\ \bibinfo {pages}
  {517--526}\BibitemShut {NoStop}%
\bibitem [{\citenamefont {Stute}\ \emph {et~al.}(2013)\citenamefont {Stute},
  \citenamefont {Casabone}, \citenamefont {Brandst{\"{a}}tter}, \citenamefont
  {Friebe}, \citenamefont {Northup},\ and\ \citenamefont
  {Blatt}}]{stute_quantum-state_2013}%
  \BibitemOpen
  \bibfield  {author} {\bibinfo {author} {\bibfnamefont {A.}~\bibnamefont
  {Stute}}, \bibinfo {author} {\bibfnamefont {B.}~\bibnamefont {Casabone}},
  \bibinfo {author} {\bibfnamefont {B.}~\bibnamefont {Brandst{\"{a}}tter}},
  \bibinfo {author} {\bibfnamefont {K.}~\bibnamefont {Friebe}}, \bibinfo
  {author} {\bibfnamefont {T.~E.}\ \bibnamefont {Northup}},\ and\ \bibinfo
  {author} {\bibfnamefont {R.}~\bibnamefont {Blatt}},\ }\bibfield  {title}
  {\bibinfo {title} {Quantum-state transfer from an ion to a photon},\ }\href
  {http://www.nature.com/articles/nphoton.2012.358} {\bibfield  {journal}
  {\bibinfo  {journal} {Nat. Photon.}\ }\textbf {\bibinfo {volume} {7}},\
  \bibinfo {pages} {219} (\bibinfo {year} {2013})}\BibitemShut {NoStop}%
\bibitem [{\citenamefont {Wright}\ \emph {et~al.}(2018)\citenamefont {Wright},
  \citenamefont {Francis-Jones}, \citenamefont {Gawith}, \citenamefont
  {Becker}, \citenamefont {Ledingham}, \citenamefont {Smith}, \citenamefont
  {Nunn}, \citenamefont {Mosley}, \citenamefont {Brecht},\ and\ \citenamefont
  {Walmsley}}]{wright_two-way_2018}%
  \BibitemOpen
  \bibfield  {author} {\bibinfo {author} {\bibfnamefont {T.~A.}\ \bibnamefont
  {Wright}}, \bibinfo {author} {\bibfnamefont {R.~J.~A.}\ \bibnamefont
  {Francis-Jones}}, \bibinfo {author} {\bibfnamefont {C.~B.~E.}\ \bibnamefont
  {Gawith}}, \bibinfo {author} {\bibfnamefont {J.~N.}\ \bibnamefont {Becker}},
  \bibinfo {author} {\bibfnamefont {P.~M.}\ \bibnamefont {Ledingham}}, \bibinfo
  {author} {\bibfnamefont {P.~G.~R.}\ \bibnamefont {Smith}}, \bibinfo {author}
  {\bibfnamefont {J.}~\bibnamefont {Nunn}}, \bibinfo {author} {\bibfnamefont
  {P.~J.}\ \bibnamefont {Mosley}}, \bibinfo {author} {\bibfnamefont
  {B.}~\bibnamefont {Brecht}},\ and\ \bibinfo {author} {\bibfnamefont {I.~A.}\
  \bibnamefont {Walmsley}},\ }\bibfield  {title} {\bibinfo {title} {Two-way
  photonic interface for linking the $\mathrm{Sr}^{+}$ transition at 422 nm to
  the telecommunication {$C$} band},\ }\href
  {https://doi.org/10.1103/PhysRevApplied.10.044012} {\bibfield  {journal}
  {\bibinfo  {journal} {Phys. Rev. Applied}\ }\textbf {\bibinfo {volume}
  {10}},\ \bibinfo {pages} {044012} (\bibinfo {year} {2018})}\BibitemShut
  {NoStop}%
\bibitem [{\citenamefont {Krutyanskiy}\ \emph {et~al.}(2019)\citenamefont
  {Krutyanskiy}, \citenamefont {Meraner}, \citenamefont {Schupp}, \citenamefont
  {Krcmarsky}, \citenamefont {Hainzer},\ and\ \citenamefont
  {Lanyon}}]{krutyanskiy_light-matter_2019}%
  \BibitemOpen
  \bibfield  {author} {\bibinfo {author} {\bibfnamefont {V.}~\bibnamefont
  {Krutyanskiy}}, \bibinfo {author} {\bibfnamefont {M.}~\bibnamefont
  {Meraner}}, \bibinfo {author} {\bibfnamefont {J.}~\bibnamefont {Schupp}},
  \bibinfo {author} {\bibfnamefont {V.}~\bibnamefont {Krcmarsky}}, \bibinfo
  {author} {\bibfnamefont {H.}~\bibnamefont {Hainzer}},\ and\ \bibinfo {author}
  {\bibfnamefont {B.~P.}\ \bibnamefont {Lanyon}},\ }\bibfield  {title}
  {\bibinfo {title} {Light-matter entanglement over 50 km of optical fibre},\
  }\href {https://doi.org/10.1038/s41534-019-0186-3} {\bibfield  {journal}
  {\bibinfo  {journal} {npj Quantum Inf.}\ }\textbf {\bibinfo {volume} {5}},\
  \bibinfo {pages} {72} (\bibinfo {year} {2019})}\BibitemShut {NoStop}%
\bibitem [{\citenamefont {Hannegan}\ \emph {et~al.}(2022)\citenamefont
  {Hannegan}, \citenamefont {Siverns},\ and\ \citenamefont
  {Quraishi}}]{hannegan_entanglement_2022}%
  \BibitemOpen
  \bibfield  {author} {\bibinfo {author} {\bibfnamefont {J.}~\bibnamefont
  {Hannegan}}, \bibinfo {author} {\bibfnamefont {J.~D.}\ \bibnamefont
  {Siverns}},\ and\ \bibinfo {author} {\bibfnamefont {Q.}~\bibnamefont
  {Quraishi}},\ }\bibfield  {title} {\bibinfo {title} {Entanglement between a
  trapped-ion qubit and a 780-nm photon via quantum frequency conversion},\
  }\href {https://doi.org/10.1103/PhysRevA.106.042441} {\bibfield  {journal}
  {\bibinfo  {journal} {Phys. Rev. A}\ }\textbf {\bibinfo {volume} {106}},\
  \bibinfo {pages} {042441} (\bibinfo {year} {2022})}\BibitemShut {NoStop}%
\bibitem [{\citenamefont {Bourdeauducq}\ \emph {et~al.}(2021)\citenamefont
  {Bourdeauducq} \emph {et~al.}}]{ARTIQ}%
  \BibitemOpen
  \bibfield  {author} {\bibinfo {author} {\bibfnamefont {S.}~\bibnamefont
  {Bourdeauducq}} \emph {et~al.},\ }\href
  {https://doi.org/10.5281/zenodo.1492176} {\bibinfo {title} {{m-labs/artiq:
  6.0 (Version 6.0)}}} (\bibinfo {year} {2021})\BibitemShut {NoStop}%
\end{thebibliography}%

\clearpage
\onecolumngrid
\begin{center}
    \textbf{\large Supplemental Material for `\Title{}'}
\end{center}
\twocolumngrid
%%%%%%%%%% Prefix a "S" to all equations, figures, tables and reset the counter %%%%%%%%%%
\setcounter{equation}{0}
\setcounter{figure}{0}
\setcounter{table}{0}
\setcounter{page}{1}
\makeatletter
\renewcommand{\theequation}{S\arabic{equation}}
\renewcommand{\thefigure}{S\arabic{figure}}
\renewcommand{\thetable}{S\arabic{table}}

\section{S1.\ Data Analysis}
\subsection{S1.1.\ Maximum Likelihood Ion-Photon Tomography}\noindent
We perform maximum likelihood estimation of the joint ion-photon density matrix, for each detector $h\in\{0\dots 3\}$, by numerically minimizing the negative logarithm of the likelihood function
\begin{multline*}
    \mathcal{L} \sim
    \prod_{i} \mathrm{Tr}\left(\hat{\Pi}_{\emptyset, i} \hat{\rho}_P\right)^{n_{\emptyset,i}}
    \prod_{k\neq h} \mathrm{Tr}\left(\hat{\Pi}_{k, i} \hat{\rho}_P\right)^{n_{k,i}}\\
    \times \prod_{j} \mathrm{Tr}\left[\left(\hat{\Xi}_{j,\mathrm{bright}}\otimes\hat{\Pi}_{h,i}\right) \hat{\rho}_{IP} \right]^{n_{h, i, \mathrm{bright}, j}}\\
    \times \mathrm{Tr}\left[\left(\hat{\Xi}_{j, \mathrm{dark}} \otimes \hat{\Pi}_{h,i}\right) \hat{\rho}_IP\right]^{n_{h, i, \mathrm{dark}, j}} \ ,
\end{multline*}
where $\hat{\Pi}$ and $\hat{\Xi}$ denote the POVMs acting on the photonic and ionic part, respectively, $n$ are the number of experimental occurrences and $\hat{\rho}_P = \mathrm{Tr}_{I}(\hat{\rho}_{IP})$ is the purely photonic part of the joint ion-photon density matrix $\hat{\rho}_{IP}$.
Index $i$ enumerates the photon polarization measurement bases, and $k$ enumerates the other detectors, for which only the likelihood of the photonic measurements contributes.
Index $j$ enumerates the ion qubit measurement bases, and dark/bright indicates the measurement outcome.

\begin{table}[hbp]
    \begin{tabular}{rlc@{\hspace{30pt}}rl}
        \cline{1-2}\cline{4-5}
        Parameter & Value && Parameter & Value \\
        \cline{1-2}\cline{4-5}
        $r_{\lambda/4}$ & $0.217 \times 2\pi$ && $\epsilon_{\mathrm{A}, H}$ & 1:12500 \\
        $r_{\lambda/2}$ & $0.449 \times 2\pi$ && $\epsilon_{\mathrm{A}, V}$ & 1:700 \\
        $t_{\mathrm{BS},H}$ & $0.5283$ && $\epsilon_{\mathrm{B}, H}$ & 1:3000 \\
        $t_{\mathrm{BS},V}$ & $0.5307$ && $\epsilon_{\mathrm{B}, V}$ & 1:1900
    \end{tabular}
    \label{tab:BellStateAnalyzer}
    \caption{Independently characterized parameters of optical elements in the photonic Bell state analyzer.
        Subscripts A and B refer to the optical paths corresponding to the two output ports of the non-polarizing beam splitter (BS).
        $H$ and $V$ refer to orthogonal linear polarization states aligned with the axis of the polarizing beam splitter.}
\end{table}

The detectors used in this experiment are embedded in a photonic Bell state analyzer with waveplates for polarization control at the input port.
The second input port is reserved for quantum networking experiments and is left open here.
The unitary map $\hat{U}_{P}$ from the polarization basis of an incoming photon to the four spatially separated, $\{H, V\}\ni q$-polarized output modes $\ket{h,q}$ is a function of waveplate retardances $r$, waveplate rotation angles relative to the fast axis $\beta$, transmission of the non-polarizing beam splitter $t_{\mathrm{BS}}$ and the polarization extinction ratios of the polarizing beam splitters $\epsilon$ (see Table~S1 for independently measured values).

The POVMs for detection of a photon on a particular detector $h$, dropping index $i$ for readability, are
\begin{align*}
    \hat{\Pi}_h &= \sum_{q\in\{H, V\}}\eta_h \hat{U}_{P}^{\dagger} \ket{h,q}\bra{h,q} \hat{U}_{P} \ , \\
    \hat{\Pi}_{\emptyset} &= \hat{1} - \sum_{k=0}^3\hat{\Pi}_k \ ,
\end{align*}
where $\eta_h$ is the overall detection efficiency for this detector.
We model the waveplates as unitary retarders.
By placing the quarter waveplate nearest to the PBS, followed by the half waveplate, a measurement along a rotated polarization basis is implemented.
For tomography, the half waveplate angle is chosen from $\{0, \pi/8\}$ and the quarter waveplate angle from $\{0, \pi/4\}$ with respect to the respective fast axis.
The projector for the ion state is
\begin{align*}
    \hat{\Xi}_s(\vartheta, \varphi) &= U_I(\vartheta, \varphi)^{\dagger} \ket{s}\bra{s} U_I \\
    \text{where} \ U_I &= \begin{bmatrix}
        \cos(\vartheta / 2) & -i \mathrm{e}^{i \varphi}\sin(\vartheta / 2) \\
        -i \mathrm{e}^{-i \varphi}\sin(\vartheta / 2) & \cos(\vartheta / 2)
    \end{bmatrix} \ ,
\end{align*}
expressed in the computational qubit basis, with $\vartheta, \varphi$ parameterizing the rotation.
For tomography, we use the settings $\vartheta=\pi /2, \varphi \in \{0, \pi/4, \pi/2, 3\pi/4\}$ and twice the setting $\vartheta=0$ to establish an over-complete set of measurement bases.
For ion-photon tomography, we exhaust all 24 measurement basis settings, collecting (typically) 500 ion-photon measurements at every step.
For simultaneous tomography on both \Sr{88} and \Ca{43}, the same measurement basis settings are used for both ions and both photons.
The collected data from the two ion-photon states are analyzed separately using the MLE method described above.

\subsection{S1.2.\ Ion-Photon Entanglement Fidelity}\noindent
To quantify the fidelity of entanglement in a two-qubit system, we follow the procedure outlined in Ref.~\cite{badziag_local_2000}.
The fidelity of a state $\hat{\rho}$ is defined by
\begin{align*}
    \mathcal{F}[\hat{\varrho}]=\max_{\hat{U}} \braket{\Psi^{+} | {\hat{U} \hat{\varrho}\hat{U}^\dagger} | \Psi^{+}} \ ,
\end{align*}
where $\ket{\Psi^{+}} = (\ket{\uparrow}\otimes\ket{\uparrow}+\ket{\downarrow}\otimes\ket{\downarrow})/\sqrt2$ is a maximally entangled two-qubit state and $\hat{U} = \hat{U}_1\otimes \hat{U}_2$ is a unitary operator.
We express a general bipartite state as
\begin{multline*}
    \hat{\varrho} = \frac14\left(\hat{1}\otimes\hat{1}+\sum_{k=1}^3 a_k \hat{\sigma}_k\otimes\hat{1} + \sum_{k=1}^3 b_k \hat{1}\otimes\hat{\sigma}_k + \right.\\
    \left.\sum_{m,n=1}^3 t_{mn}\hat{\sigma}_m\otimes\hat{\sigma}_n\right)  \ ,
\end{multline*}
and calculate the corresponding fidelity,
\begin{align*}
    \mathcal{F} = \max_{\hat{U}} \left( 1 + \tilde{t}_{11} - \tilde{t}_{22} + \tilde{t}_{33} \right)/4 \ ,
\end{align*}
where $\sum_{m,n=1}^3\tilde{t}_{mn}\hat{\sigma}_m\otimes\hat{\sigma}_n = \sum_{m,n=1}^3t_{mn} (\hat{U}_1\otimes \hat{U}_2)\hat{\sigma}_m\otimes\hat{\sigma}_n (\hat{U}_1\otimes \hat{U}_2)^\dagger$.
As $\mathrm{SU(2)}\cong\mathrm{SO(3)}/\{\pm 1\}$, we can write
\begin{align}
    T &= O_1^\dagger \tilde{T} O_2 \label{eq:IsomorphismSU2SO3}
    \intertext{with $O_1, O_2 \in {\rm SO(3)}$ and $T,\tilde{T}$ are real $3\times 3$ matrices with entries $t_{mn},\tilde{t}_{mn}$.
        It is possible to choose $O_1$ and $O_2$ such that $\tilde{T}={\rm diag}(\tilde{t}_{11},\tilde{t}_{22},\tilde{t}_{33})$ with $|\tilde{t}_{11}|\geq |\tilde{t}_{22}| \geq |\tilde{t}_{33}|$.
        In practice, this is done using the singular value decomposition}
    T&=V\Sigma W^\dagger \ , \label{eq:SVD}
\end{align}
where $V,W$ are orthogonal $3\times 3$ matrices and $\Sigma={\rm diag}(s_1,s_2,s_3)$ with conventional ordering of singular values $s_{1}\geq s_2 \geq s_3\geq 0$.
The difference between the decomposition \eqref{eq:IsomorphismSU2SO3} and \eqref{eq:SVD} is that $\det(\Sigma) = \pm\det(T)$, whereas $\det(\tilde{T})=\det(T)$, which expresses that, unlike $O_1$ and $O_2$, $V$ and $W$ are not proper rotations.
Nevertheless, the sign difference is easily included:
\begin{align*}
    \mathcal{F} = \left\{ 1 + s_1 + s_2 - s_3 \,{\rm sign}[\det(T)] \right\}/4 \ , \label{eq:FullyEntangledFraction}
\end{align*}
which maximizes the entangled fraction.
States with $\det(T)\geq 0$ are classical.

\subsection{S1.3.\ Nonparametric Bootstrapping}\noindent
Each data set can be resampled according to inherent statistical properties, in order to generate new data sets, which can be analyzed in the same way as the measured data.
The bootstrapped results approximate the measured result on average, but with a spread indicating how sensitive the analysis is to statistical fluctuations in the input data.
For the ion-photon experiments presented here, the number of excitation attempts $n$ is Poisson distributed, photons are multinomially distributed into the four detectors, and the thresholded ion-fluorescence readout follows binomial statistics.
The underlying probabilities for these distributions are estimated from the measured data set in frequentist manner.
To bootstrap process tomography measurements on two ions, combined ion-fluorescence readout results are resampled according to the multinomial distribution with four outcomes, using the experimentally observed frequentist probabilities as prior information.

\section{S2.\ Experimental Methods}
\subsection{S2.1.\ Cooling}\noindent
Each experimental sequence begins with Doppler cooling on both ions for \SI{300}{\micro\second}.
Next, a sequence of \num{30} red-sideband Raman pulses with durations linearly increasing from \SI{40}{\micro\second} to \SI{80}{\micro\second}, interleaved with optical pumping into the \Ca{43} logic qubit, pre-cool the \ac{OOP} axial mode at $\omega_\mathrm{oop}/(2\pi)=\OopModeFrequency{}$.
Finally, alternating pulses of electromagnetically-induced transparency-cooling target the two axial modes on \Sr{88} for a total of \SI{1.1}{\milli\second}.

\subsection{S2.2.\ Transfer between Ca-43 Logic and Memory Qubits}\noindent
The memory qubit, $\ket{\hfslevshort{4}{0}} \leftrightarrow \ket{\hfslevshort{3}{0}}$, is magnetic field-insensitive at low magnetic field, but -- unlike the logic qubit, $\ket{\hfslevshort{4}{4}} \leftrightarrow \ket{\hfslevshort{3}{3}}$ -- cannot be prepared using optical pumping, and does not experience the light shift force required for our gate scheme.
Therefore, a sequence of Raman $\pi$ pulses is used to coherently transfer population between these qubits.
First, population from $\ket{\hfslevshort{3}{3}}$ is transferred into $\ket{\hfslevshort{3}{0}}$, and subsequently population from $\ket{\hfslevshort{4}{4}}$ is transferred into $\ket{\hfslevshort{3}{1}}$.
At this point, a complication arises as the $\ket{\hfslevshort{4}{1}} \leftrightarrow \ket{\hfslevshort{3}{0}}$ and the $\ket{\hfslevshort{4}{0}} \leftrightarrow \ket{\hfslevshort{3}{1}}$ transitions are only separated by $\Delta f \approx\SI{15}{\kilo\hertz}$ at the \SI{0.5}{\milli\tesla} magnetic field.
To coherently reverse the effect of off-resonant excitation on the $\ket{\hfslevshort{4}{1}} \leftrightarrow \ket{\hfslevshort{3}{0}}$ transition, we apply two $\pi/2$ pulses on the $\ket{\hfslevshort{4}{0}} \leftrightarrow \ket{\hfslevshort{3}{1}}$ transition with a delay of $\SI{157}{\micro\second}\sim 2/\Delta f$ in between.
%todo process tomo alice_457983

\subsection{S2.3.\ Errors in $\sigma_z \otimes \sigma_z$ Gate}\noindent
%todo process tomo 13-08-2022, ~23:00
We identify the leading error mechanism to be heating.
We include heating on both axial modes of motion in a master equation approach by adding the Lindblad collapse operators $\sqrt{\dot{\bar{n}}} a$ and $\sqrt{\dot{\bar{n}}} a^\dagger$ per mode, with the corresponding heating rate $\dot{\bar{n}}$, creation operator $a^\dagger$ and annihilation operator $a$.
In a mixed-species, two-ion crystal, the axial normal modes of motion are found at frequencies
\begin{equation*}
    \omega_{\substack{\mathrm{oop} \\ \mathrm{ip}}}^2 = \frac{\left(1 + \mu \pm \sqrt{1 + (\mu - 1) \mu}\right) \omega_1^2}{\mu} \ ,
\end{equation*}
where $\mu=m_2 / m_1$ is the mass ratio of the two ions, and $\omega_1$ is the axial frequency of a single ion of mass $m_1$ in the same trap.
For \Sr{88} and \Ca{43}, the ratio $\omega_\mathrm{oop}/\omega_\mathrm{ip} \approx 1.94486$ is close to $2$, resulting in unavoidable off-resonant excitation of the second harmonic of the \ac{IP} mode when addressing the \ac{OOP} mode.
In the absence of heating, a gate detuning can always be found which leaves both modes disentangled from the ions.
In our system, however, a high \ac{IP} mode heating rate of $\dot{\bar{n}}_\mathrm{ip}\simeq\IPHeatingRate{}$ causes an error, which is amplified by the high initial temperature of this mode.
We simulate the system of two ions coupled by the two axial motional modes to quantify the infidelity due to this effect.
We find that the contribution from the second harmonic of the \ac{IP} mode is of similar magnitude (\ish\num{0.02}) as the direct contribution from the $\dot{\bar{n}}_\mathrm{oop}\simeq\OOPHeatingRate{}$ heating rate of the \ac{OOP} gate mode.

\end{document}